\documentclass[fleqn,usenatbib]{mnras}

\usepackage{newtxtext,newtxmath}
 
\usepackage[T1]{fontenc}

\DeclareRobustCommand{\VAN}[3]{#2}
\let\VANthebibliography\thebibliography
\def\thebibliography{\DeclareRobustCommand{\VAN}[3]{##3}\VANthebibliography}

\usepackage{graphicx}	
\usepackage{amsmath}	
\usepackage{amssymb}	
\usepackage{tabularx}
\usepackage{multirow}
\usepackage{makecell}
\usepackage{mathtools}
\usepackage{dblfloatfix}
\usepackage{tablefootnote}
\usepackage{threeparttable}
\usepackage{upgreek}
\usepackage{caption}

\title[The Peculiar Mini-Halo in A\,3558]{The Peculiar Mini-Halo in the Shapley Supercluster Member Abell 3558}

\author[K. S. Trehaeven et al.]{K. S. Trehaeven$^{1,2}$\thanks{E-mail: ktrehaeven@gmail.com},
	T. Venturi$^{2,1}$,
	O. Smirnov$^{1,3,2}$,
	G. Di Gennaro$^{2}$,
        M. Rossetti$^{4}$,
	S. Giacintucci$^{5}$,
	\newauthor{
		S. Bardelli$^{6}$,
		P. Merluzzi$^{7}$,
            S. P. Sikhosana$^{8,9}$,
            D. Dallacasa$^{10,2}$,
            R. Kale$^{11}$},
            K. Knowles$^{1,3,9}$
	\\
    \\
	$^{1}$Centre for Radio Astronomy Techniques and Technologies (RATT), Department of Physics and Electronics, Rhodes University, Makhanda 6140, South Africa\\
	$^{2}$INAF – Istituto di Radioastronomia, via P. Gobetti 101, 40129 Bologna, Italy \\
	$^{3}$South African Radio Astronomy Observatory, Cape Town 7700, South Africa\\
    $^{4}$INAF - Istituto di Astrofisica Spaziale e Fisica Cosmica di Milano, via A. Corti 12, 20133 Milano, Italy\\
	$^{5}$Naval Research Laboratory, 4555 Overlook Avenue SW, Code 7213, Washington, DC 20375, USA\\
        $^{6}$INAF - Osservatorio di Astrofisica e Scienza dello Spazio di Bologna, via Gobetti 93/3, 40129 Bologna, Italy\\
        $^{7}$INAF - Osservatorio Astronomico di Capodimonte, Salita Moiariello 16 80134 Napoli, Italy\\
        $^{8}$Astrophysics Research Centre, University of KwaZulu-Natal, Durban, 3696, South Africa\\
        $^{9}$School of Mathematics, Statistics \& Computer Science, University of KwaZulu-Natal, Westville Campus, Durban 4041, South Africa\\
        $^{10}$Dipartimento di Fisica e Astronomia, Università degli Studi di Bologna, via Gobetti 93/2, I-40129, Bologna, Italy\\
        $^{11}$National Centre for Radio Astrophysics, Tata Institute of Fundamental Research, S. P. Pune University, Ganeshkhind, Pune 411007, India}
\date{Accepted 2025 July 04. Received 2025 July 01; in original form 2025 March 06}

\pubyear{2025}

\begin{document}
	\maketitle
	
	\begin{abstract}

        We present a multi-band study of the diffuse emission in the galaxy cluster Abell 3558, located in the core of the Shapley Supercluster. Using new MeerKAT UHF-Band and uGMRT Band-3 observations, and published MeerKAT L-band and ASKAP 887 MHz data, we perform a detailed analysis of the diffuse emission in the cluster centre. We complement with {\it XMM-Newton} X-ray information for a thorough study of the connection between the thermal and non-thermal properties of the cluster. We find that the diffuse radio emission in the cluster centre is more extended than published earlier, with a previously undetected extension spanning 100 kpc towards the north beyond the innermost cold front, increasing the total size of the emission to 550 kpc, and shows a clear spatial correlation with the X-ray features. The overall radio spectrum is steep ($\alpha_{\rm 400\,MHz}^{\rm 1569\,MHz}=1.18\pm0.10$), with local fluctuations which show several connections with the X-ray surface brightness, cold fronts, and residual emission. The point-to-point correlation between the radio and X-ray surface brightness is sub-linear, and steepens with increasing frequency. We discuss the classification of the diffuse emission considering its overall properties, those of the ICM, and the existing scaling laws between the radio and X-ray quantities in galaxy clusters. We conclude that it is a mini-halo, powered by turbulent (re)-acceleration induced by sloshing motions within the cluster region delimited by the cold fronts, and it supports the picture of a known minor merger between A\,3558 and the group SC\,1327--312 with mass ratio 5:1.

	\end{abstract}
	
	
	
	\section{Introduction}

	Superclusters are the most massive gravitationally bound structures in the Universe (M $\sim10^{16}$M$_{\odot}$). They form by hierarchical gravitational collapse from initial density perturbations \citep[e.g.,][]{2003A&A...405..425E,2003A&A...410..425E}. They contain clusters and groups over a wide range of masses, from the most massive clusters ($\sim10^{15}$M$_{\odot}$) to small groups ($\sim10^{13}$M$_{\odot}$), as well as much less dense regions connecting them (filaments). During the collapse phase, superclusters are host to a broad range of dynamical processes, from the relatively rare major merger events involving massive clusters (i.e. $\ge 6\times10^{14}$M$_{\odot}$) with mass ratios close to unity, to the much more common but less energetic minor merger events where the accreted mass is that of a group or small subcluster with mass ratios well below $M_2:M_1\sim1:4$ \citep[e.g.,][and references therein]{2016A&A...593A..81C,2023A&A...677A.169D}.
	
	The gravitational energy released in the cluster volume during major and minor mergers leaves important signatures in the properties of the hot (T$\sim10^7 - 10^8$K) intracluster gas which permeates galaxy clusters, such as shocks and cold fronts \citep{2007PhR...443....1M}, and affects their non-thermal properties \citep{2014IJMPD..2330007B}. The close connection between the properties of the radio emission in galaxy clusters and their dynamical state is now an established result \citep{2010ApJ...721L..82C,2021A&A...647A..51C}. In particular, merger activity and accretion processes induce turbulence and shocks. This energy then cascades down to the micro-physical level to produce particle (re)-acceleration and magnetic field amplification, causing Mpc-scale synchrotron steep-spectrum\footnote{Spectral index, $\alpha$, typically $\sim1.3$ for flux density S $\propto\nu^{-\alpha}$ where $\nu$ is the observing frequency.} radio sources of $\sim\upmu$Jy/arcsec$^{2}$ surface brightness at GHz frequencies. These are giant radio halos and relics. The reader is referred to \citet{2019SSRv..215...16V} for an observational overview, and \citet{2014IJMPD..2330007B} for a theoretical review. On a slightly smaller scale, mainly restricted to the central regions of the cluster \citep[up to a few hundred kpc or 0.2R$_{500},$\footnote{R$_{500}$ is the radius that encloses a mean overdensity of 500 with respect to the critical density at the cluster redshift.}][]{2017ApJ...841...71G}, radio mini-halos are found mainly in dynamically relaxed systems hosting cool-cores, where no major merger event has occurred \citep[e.g.,][]{2008ApJ...675L...9M}. Their origins are still under debate, but the preferred mechanism is turbulent re-acceleration induced by gas sloshing and/or AGN feedback processes \citep{2013ApJ...762...78Z,2020MNRAS.499.2934R,2022MNRAS.512.4210R}. However, a possible alternative is due to a continuous injection of cosmic ray electrons (CRe) produced during hadronic collisions between cosmic ray protons (CRp) and thermal protons \citep[][and references therein]{2015ApJ...801..146Z}.
    
The observed correlations between the global properties of the radio, X-ray emission and cluster mass \citep[e.g.,][]{2001ApJ...553L..15B,2006MNRAS.369.1577C,2013ApJ...777..141C,2008Natur.455..944B,2008A&A...484..327V,2015A&A...579A..92K,2019ApJ...880...70G,2021A&A...647A..51C} as well as the spatial connection between the thermal properties of the X-ray gas (extent, brightness, temperature and other thermodynamic quantities) and the radio emission \citep[extent, brightness, spectral index, e.g.,][]{2001A&A...376..803G,2005A&A...440..867G,2023MNRAS.524.6052R} are a further indication that the accretion processes are the common origin for the features observed in the radio and X-ray bands.

Until recently, our studies of cluster mergers and their observable footprints in the radio band have been limited to intermediate- to high-mass systems \citep[see][for a recent statistical investigation]{2021A&A...647A..51C,2023A&A...680A..30C}. This is simply due to the steep relation between the cluster mass (or X-ray luminosity) and the radio power for radio halos \citep{2021A&A...647A..51C}, relics \citep{2023A&A...680A..31J} and mini-halos \citep{2019ApJ...880...70G}. However, the improved sensitivity at high angular resolutions of the current generation of radio astronomical facilities, namely the LOw Frequency ARray \citep[LOFAR,][]{2013A&A...556A...2V}, the Australian Square Kilometre Array Pathfinder \citep[ASKAP,][]{2021PASA...38....9H}, the \textit{Meer} Karoo Array Telescope \citep[MeerKAT,][]{2016mks..confE...1J}, and the Upgraded Giant Metrewave Radio Telescope \citep[uGMRT,][]{2017CSci..113..707G}, have paved the way to the exploration of new parameter spaces and revealed new phenomena that have been inaccessible thus far. A couple of remarkable examples are the giant radio halos detected beyond the previously well-known mini-halos in A\,2142 \citep{2017A&A...603A.125V,2023A&A...678A.133B} and in the Perseus cluster \citep{2024A&A...692A..12V}. These show the complexity of merging processes and the limitations of our sharp classifications of the radio emission. Further examples are the faint radio bridges of emission connecting galaxy clusters and groups, such as the Shapley bridge \citep{2022A&A...660A..81V}, that in the Coma cluster \citep{2021ApJ...907...32B}, Abell 1758N-S \citep{2020MNRAS.499L..11B} and the Abell 399-401 pair \citep{2019Sci...364..981G,2024A&A...685L..10P}.

The focus of this paper is the diffuse radio emission at the centre of the massive cluster Abell 3558 (hereafter A\,3558). This cluster is located in the core region of the Shapley Supercluster (hereinafter SSc), the most massive supercluster in the local Universe \citep[][M$_{\rm tot}=5\times10^{16}$ h$^{-1}$ M$_{\odot}$]{2003A&A...405..425E} at a mean redshift of $\langle z \rangle \sim 0.048$. The SSc includes high- to intermediate-mass clusters as well as small groups and filaments in different dynamical stages over a 12$\times$30 deg$^2$ region \citep{2006A&A...447..133P,2020A&A...638A..27Q}, and is the perfect environment to study the effects of major and minor mergers from the supercluster/intercluster scale all the way down to that of individual galaxies.

MeerKAT L-band and ASKAP 887 MHz radio observations of the SSc led to the detection of diffuse emission both on the cluster scale in A\,3562 \citep[][and references therein]{2022ApJ...934...49G} and A3558, and on the intercluster scale with the detection of a radio bridge connecting A\,3562 and SC\,1329--313 which is coincident with the X-ray emission detected by {\it XMM-Newton} \citep{2022A&A...660A..81V}. The multi-band coverage suggests that the core of the SSc is undergoing minor merger events \citep[see for instance][for a discussion on the region between A\,3562 and SC\,1329--313]{2002A&A...382...17B,2004ApJ...611..811F,2022ApJ...934...49G,2022A&A...660A..81V}.

To further explore the ongoing dynamics of the SSc, here we present a detailed study of the diffuse radio emission at the centre of A\,3558 using deep observations carried out with MeerKAT UHF-band (816 MHz) and uGMRT Band3 (400 MHz), complemented by the MeerKAT L-band (1283 MHz) and ASKAP (887 MHz) data reported in \citet{2022A&A...660A..81V}. The paper is organised as follows: Section \ref{sec:a3558} briefly describes the A3558 environment; Section \ref{sec:data_red} describes the radio and {\it XMM-Newton} data reductions; in Section \ref{sec:continuum_imgs} we present the radio and X-ray images; in Section \ref{sec:analysis} we report our global and local radio analysis; in Section \ref{sec:correlations} we derive the radio/X-ray correlations; in Section \ref{sec:discussion} we classify the nature of the diffuse emission and discuss its origin in the framework of the ongoing minor merger in the SSc and finally in Section \ref{sec:conclusions} we summarize our results and provide future prospects.

Throughout the paper we assume a $\Lambda$CDM cosmology, with H$_0$ = 70 km/s and $\Omega_{\rm M}$ = 0.3. At the average redshift of the SSc core, $\langle z \rangle = 0.048$, 1$^{\prime\prime}$ = 0.928 kpc. We adopt the spectral index convention of S $\propto \nu^{-\alpha}$.

\section{The galaxy cluster Abell 3558}
\label{sec:a3558}

A\,3558 has been extensively observed in the X-ray, optical and radio bands. It lies at the centre of a chain of three clusters and two groups, from west to east: A\,3556 - A\,3558 - SC\,1327--312 - SC\,1329--313 - A\,3562). This nearly East-West continuous structure of interacting and merging clusters \citep[e.g.,][]{2015MNRAS.446..803M,2018MNRAS.481.1055H} spans $\sim3^{\circ}$ on the sky and is permeated by low surface brightness X-ray gas \citep{1997MNRAS.289..787E,2004ApJ...611..811F,2014A&A...571A..16P,2016MNRAS.460.3345M}. 

A\,3558 itself has redshift $z = 0.0474$ with dynamical mass $M_{\rm dyn} = 14.8\pm1.4 \times10^{14}$ $M_{\odot}$ \citep{2020MNRAS.497...52H}, a bolometric X-ray luminosity of $L_{\rm X}^{\rm bol} \sim$ 7$\times10^{44}$ ergs/s \citep{2005A&A...444..387D} and average recession velocity $\langle V_H \rangle = 14500\pm45$ km/s \citep{2018MNRAS.481.1055H}. The BCG and X-ray peak are coincident but offset from the Sunyaev Zeldovich (SZ) centre \citep{2014A&A...571A..29P} by $\sim$16$^{\prime\prime}$ (15 kpc).

\begin{figure}
	\centering
	\includegraphics[width=\columnwidth]{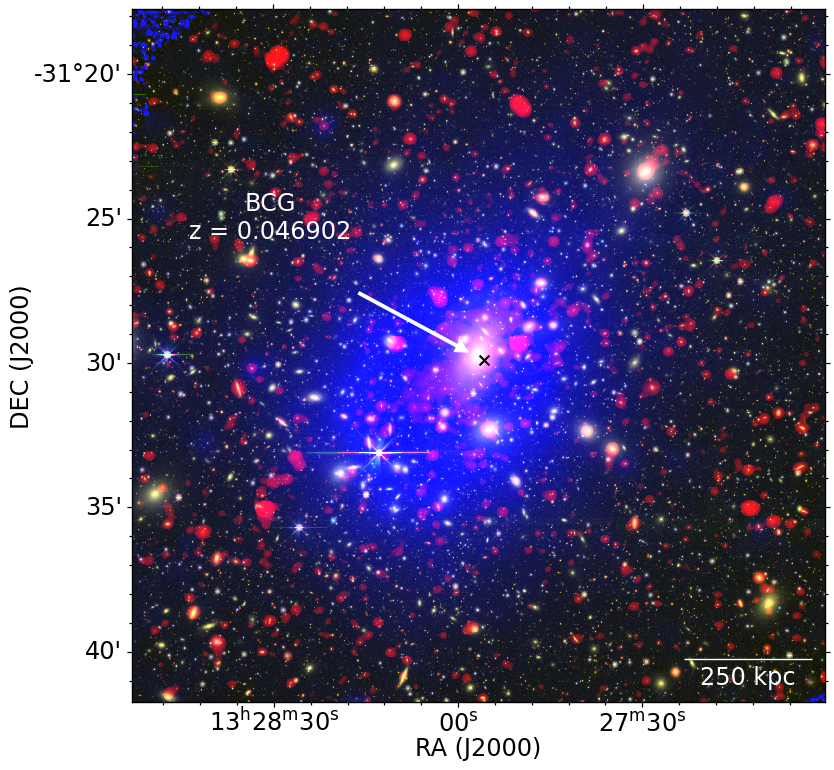}
	\caption{Composite image of the A\,3558 environment. The background is an RGB colour image of the DESI Legacy Survey DR10 in the g,r,z-bands. Overlaid in blue is the X-ray \textit{XMM-Newton} 0.7-1.2 keV image and in red the Native MeerKAT UHF-band radio image at an angular resolution of $\sim$13$^{\prime\prime}$. The BCG is labelled with its redshift. The SZ cluster peak is indicated by a black cross.}
	\label{fig:radio_xray_optical_overlay}
\end{figure}

\citet{2007A&A...463..839R} studied the thermal properties of A\,3558 in detail using deep \textit{XMM-Newton} data. They found that the intracluster medium (ICM) has mixed characteristics: a peaked metal abundance and brightness distribution typical of relaxed cool-core clusters and an asymmetric temperature distribution typical of merging clusters. Furthermore, they found a sharp discontinuity in brightness, temperature, and density at 113$^{{\prime\prime}}$ (105 kpc) NW of the cluster centre, which they attribute to a cold front. This cold front is moving NW and is believed to be driven by core oscillations or \textit{sloshing} caused by past interactions, possibly with the group SC\,1327--312, which lies $\sim22^{\prime}$ (1.2 Mpc) south-east of A\,3558 with a mass ratio $\sim1:5$. This minor merger would have mildly disturbed the cluster gravitational potential well but was likely not strong enough to destroy its structure. The sloshing oscillations likely caused the core’s dense gas to shift and form the current cold front, and hydrodynamic processes have formed a tail of low-entropy, high-pressure, metal-rich cool-core gas extending behind it. \citet{2023MNRAS.526L.124M}, using similar analyses, showed that A\,3558 hosts two large-scale cold fronts in an outwardly spiralling pattern, one at 9.7$^{\prime}$ ($\sim$540 kpc) SE and the other 20.7$^{\prime}$ ($\sim$1.2 Mpc) NW from the surface brightness peak. These cold fronts highlight the dynamic nature of the cluster ICM and suggest gas sloshing throughout the cluster volume. 

The A\,3558 complex was also included in a comprehensive optical/near-infrared survey of the entire SSc, namely the Shapley Supercluster Survey \citep[ShaSS,][]{2015MNRAS.446..803M,2016MNRAS.460.3345M,2018MNRAS.481.1055H}. In particular, to study the impact of cluster-scale mass assembly on galaxy evolution, through optical (\textit{ugri}-bands) with the ESO-VLT Survey Telescope, a weak-lensing map of the cluster core was derived, finding that the dark matter distribution is slightly offset with respect to the X-ray emission and optical galaxy density, attesting to the complex dynamical state of the core of the SSc \citep{2015MNRAS.446..803M}.

Regarding the radio regime, the galaxy population of A\,3558 has been studied by \citet{2000MNRAS.314..594V} via an Australia Telescope Compact Array (ATCA) 22 cm survey. They find that the radio source counts are consistent with the background field despite the higher optical density. Thereafter, \citet{2004A&A...419...71G} and \citet{2005AJ....130.2541M} via Very Large Array (VLA) 21 cm observations found that the radio luminosity function for early-type galaxies is lower than expected and that the cluster lacks significant radio-detected star-forming galaxies, implying that its dense environment and past merger events may have quenched star formation, but also enhanced AGN activity in its most massive galaxies. \citet{2018A&A...620A..25D} studied the Brightest Cluster Galaxies (BCGs) and tailed radio galaxies of the clusters in the SSc using GMRT 235, 325 and 610 MHz and the VLA 8.46 GHz, along with data from the TIFR GMRT Sky Survey (TGSS) at 150 MHz, the Sydney University Molonglo Sky Survey (SUMSS) at 843 MHz and ATCA at 1380, 1400, 2380, and 4790 MHz. They find that the BCG in A\,3558 is a faint compact source with a concave spectrum that is relatively steep, suggesting that this source has already aged and will not evolve beyond its present size (which we measure to be $\sim$24$^{\prime\prime}$ or 22 kpc).
However, significant large-scale diffuse emission in A\,3558 has only been detected with the current generation of radio interferometers, namely MeerKAT and ASKAP \citep{2022A&A...660A..81V}. Figure \ref{fig:radio_xray_optical_overlay} shows the central environment of A\,3558 with an optical DESI Legacy Survey DR10 \citep{2019AJ....157..168D} grz-band image, with the XMM-Newton 0.7-1.2 keV image overlaid in blue and the MeerKAT UHF-band image in red.

\section{Data reduction}
\label{sec:data_red}

We used a combination of new and previously published data to study the diffuse radio emission in A\,3558. We made use of the MeerKAT L-band visibilities from \citet{2022A&A...660A..81V}, which we reprocessed, and their final ASKAP 887 MHz image. Table \ref{tab:Table 1} summarises the observation details of the new data. The reader is referred to \citet{2022A&A...660A..81V} for the observational details of the MeerKAT L-band and ASKAP 887 MHz data. Details of the reduction of each dataset are as follows. The reader is referred to \citet{2022A&A...660A..81V} for details of the ASKAP reduction. Since we did not reprocess the ASKAP data, and point source subtraction was performed in the image-plane, we adopt a conservative fluxscale uncertainty of 20\% for all measurements on this data. We also note that all data have pointing centres on or very near the cluster BCG so primary beam correction was not necessary.

\begin{table}
	\caption{Observation details of the new data used in this study.}
	\centering
	\begin{tabular}{ll}
		\hline
		MeerKAT UHF \\
		\hline
		RA, DEC (J2000) & 13h27m54.8s, -31$^{\circ}$29$^{\prime}$32.0$^{\prime\prime}$  \\
		On-source integration & 9 hrs \\
		Frequency range & 544-1088 MHz \\
		Central frequency & 816 MHz \\
		Number of channels & 4096 \\
		Observation date & 31 March - 01 April 2022 \\
		\hline
		uGMRT Band3 \\
		\hline
		RA, DEC (J2000) & 13h27m58.0s, -31$^{\circ}$30$^{\prime}$30.5$^{\prime\prime}$  \\
		On-source integration & 5.5 hrs\\
		Frequency range & 300-500 MHz \\
		Central frequency & 400 MHz \\
		Number of channels & 8192 \\
		Observation date & 06 June 2022 \\
		\hline
	\end{tabular}
	\label{tab:Table 1}
\end{table}

\subsection{MeerKAT}

The unpublished MeerKAT data used here were downloaded from the SARAO archive (Schedule Block: 20220329-0020) and were observed as part of the observatory's UHF-band science verification program. The observations were conducted on the night of March 31 2022, totalling nine hours on-target in UHF-band - at a central frequency 816 MHz with 544 MHz total bandwidth - with full wideband 4096 channelisation and full Stokes at an 8-second dump rate. The standard MeerKAT continuum observation strategy was implemented, with J\,1939--6342 as the primary calibrator and J\,1311--2216 as the secondary calibrator, respectively.

\begin{table*}
    \begin{threeparttable}
	\caption{Properties of the new images used in this study.}
	\centering
	\begin{tabular}{ccccccccc}
	\hline
	Telescope & $\Delta \nu$ & $\nu$ & Type & Robust & UV-range & Taper & Noise & Beam \\
        & (MHz) & (MHz) & & & (k$\lambda$) & ($^{\prime\prime}$) & ($\mu$Jy/beam) & ($^{\prime\prime}\times^{\prime\prime}, ^{\circ}$)\\
	\hline
	\multirow{3}{*}{uGMRT} & \multirow{3}{*}{300-500} & \multirow{3}{*}{400} & Native & 0 & - & - & 29.4 & 9.6$\times$6.0, 173 \\
	& & & HR & 0 & > 8.4 & - & 26.3 & 7.7$\times$4.7, 2 \\
	& & & SRC-SUB & 0.5 & - & 23 & 95.0 & 30$\times$30, 0.0 \\
	\hline
	\multirow{3}{*}{MeerKAT} & \multirow{3}{*}{544-1088} & \multirow{3}{*}{816} & Native & 0 & - & - & 14.5 & 13.4$\times$12.2, 99 \\
	& & & HR & -1.5 & > 8.4 & - & 18.5 & 6.6$\times$6.4, 44 \\
	& & & SRC-SUB & 0 & 0.2-18.3 & 23 & 32.0 & 30$\times$30 \\
	\hline
	\multirow{3}{*}{MeerKAT} & \multirow{3}{*}{856-1712} & \multirow{3}{*}{1283} & Native & 0 & - & - & 5.7 & 7.9$\times$7.0, 171 \\
	& & & HR & 0 & > 8.4 & - & 6.9 & 6.9$\times$6.2, 22 \\
	& & & SRC-SUB & 0 & 0.2-18.3 & 26.5 & 17.0 & 30$\times$30 \\
	\hline
        \label{tab:Table 2}
	\end{tabular}
	\begin{tablenotes}
	   \item \textbf{Notes.} \textit{Native} refers to the final self-calibrated images, \textit{HR} refers to the high-resolution images from which point sources were modelled, \textit{SRC-SUB} refers to the compact source subtracted images. All SRC-SUB images were convolved with a Gaussian kernel to achieve the reported circular beam size.
	\end{tablenotes}
    \end{threeparttable}
\end{table*}

We used the \texttt{CARACal} pipeline \citep{2020ascl.soft06014J} to automate the transfer calibration process for both the L- and UHF-band data. Within this pipeline, the standard \texttt{CASA flagdata} task and \texttt{Tricolour}\footnote{flagging strategy: \url{https://github.com/caracal-pipeline/caracal/blob/master/caracal/data/meerkat_files/gorbachev.yaml}} \citep{2022ASPC..532..541H} were used to flag the data before calibration. Thereafter, calibration solutions were derived from the primary calibrator against its known local sky model. This was repeated after mild automated flagging. The bandpass corrections were applied to the secondary on-the-fly while its solutions were derived and scaled to the primary's flux density scale. The primary bandpass solutions and the secondary delay and flux-scaled gain solutions were applied to the target data, which were then frequency-averaged over four channels to save on storage and processing time. Finally, the cross-calibrated target data was flagged similarly to the calibrator data. 

Self-calibration was automated using the Python3 scripting framework \texttt{Stimela2} \citep{2024arXiv241210080S}, which allowed for greater flexibility than the current \texttt{CARACal} self-calibration workflow. We used the calibration software \texttt{QuartiCal} \citep{2024arXiv241210072K} to derive direction-independent solutions. For imaging, we used \texttt{WSClean v2.10}   to image at \texttt{robust 0} and multiscale cleaning with eight subbands each in both observing bands. The UHF-band cross-calibrated target field showed a large background diffraction pattern of smoothly varying positive and negative excess noise all throughout the primary beam, most likely due to persistent low-level RFI on the shortest baselines. The L-band data had a similar pattern but was much less severe. We used \texttt{breizorro}\footnote{\url{https://github.com/ratt-ru/breizorro}} to make a deep mask of the field and then re-imaged with this mask to ensure no artefacts were carried into self-calibration. \texttt{QuartiCal} read these skymodels and derived and applied direction-independent delay (K) and complex diagonal gain (G) solutions simultaneously. For both bands, the K-term was solved per integration over the entire bandwidth and the G-term was conservatively solved every three minutes in four equal subbands. After much experimentation, we found that to eliminate the large-scale diffraction patterns, we needed to enable inside \texttt{QuartiCal} a small inner uv-cut (all data > 50 m in UHF-band and > 150 m in L-band) and per baseline MAD (median absolute deviation) flagging. Some low-level fluctuations remained in both bands but did not affect our science goals. 

The final self-calibrated (\textit{Native} in Table \ref{tab:Table 2}) UHF-band images have a local noise of 14.5 $\mu$Jy/beam at a resolution of (13.4$^{\prime\prime}\times$12.2$^{\prime\prime}$, 99$^{\circ}$), and the L-band images have noise 5.7 $\mu$Jy/beam at a resolution of (7.9$^{\prime\prime}\times$7.0$^{\prime\prime}$, 171$^{\circ}$).



The MeerKAT L- and UHF-bands have an overlapping frequency range from 856-1088 MHz. To estimate the flux scale uncertainty in the UHF-band, we re-imaged the final calibrated L and UHF target fields within the frequency range 890-910 MHz (central frequency of 900 MHz). Then, \texttt{PyBDSF} \citep{2015ascl.soft02007M} was used to identify and measure point source flux densities within the primary beam across the two data sets. We calculated the mean percentage difference between the two datasets to be 8-10\% and thus conservatively adopted a 10\% fluxscale uncertainty on the UHF data.


\subsection{uGMRT}

The uGMRT Band-3 data (300-500 MHz, with 400 MHz as central frequency) were part of a survey covering the core of the SSc with five pointings \citep[Proposal Id: 42\_019, see][]{2024MNRAS.533.1394M}. The observations were conducted on June 5th 2022 for a total time of seven hours, including times on the calibrators (i.e. 3C\,286 and J\,1311--222 as primary and secondary calibrators, respectively). Data were recorded in 8192 frequency channels, with an 8-second dump rate in full Stokes mode.

We processed the data using the Source Peeling and Atmospheric Modeling \citep[\texttt{SPAM,}][]{2014ASInC..13..469I} software. Since both narrow- and wide-band correlators were used, we first reduced the narrow-band data and then the wide-band data, using the narrow-band image as a skymodel. The wide-band data were split into six subbands of 33.3 MHz bandwidth each. These were calibrated independently and then imaged together using \texttt{WSClean~v2.10} at the common frequency of 400 MHz. The final wideband image has a noise of 29.4 $\mu$Jy/beam at a resolution of (9.6$^{\prime\prime}\times$6.0$^{\prime\prime}$, 173$^{\circ}$). Final systematic uncertainties due to residual amplitude errors are set to 8\% \citep{2004ApJ...612..974C}.

\subsection{Compact source subtraction}

\begin{figure}
	\centering
	\includegraphics[width=\columnwidth]{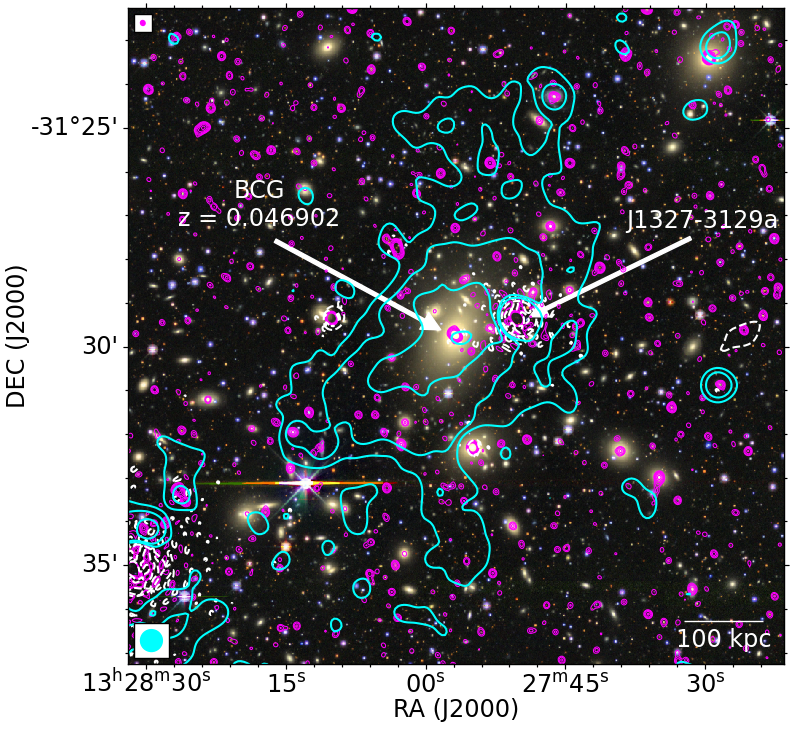}
	\caption{A\,3558 optical/radio overlay. The colormap is the same DESI image from Figure \ref{fig:radio_xray_optical_overlay}. The magenta contours show our high-resolution (HR as listed in Table \ref{tab:Table 2}) MeerKAT L-band 1283 MHz emission at a 6.9$^{\prime\prime}\times$6.2$^{\prime\prime}, 21.6^{\circ}$ resolution (shown in the top left corner) with noise 6.9$\mu$Jy/beam, starting at $3\sigma$ and increasing by a factor of two. The cyan contours show the low-resolution point source subtracted (SRC-SUB in Table \ref{tab:Table 2}) MeerKAT L-band 1283 MHz diffuse emission studied in this work at a 30$^{\prime\prime}$ resolution (bottom left) with $\sigma$=17$\mu$Jy/beam, starting at $3\sigma$ and increasing by a factor of two. The dashed white contours show the -$3\sigma$ level at both high and low resolutions. The cluster BCG is labelled along with a powerful embedded source, J\,1327--3129a.} 
\label{fig:radio_optical_overlay}
\end{figure}

The diffuse emission in A\,3558 hosts many compact sources, a few of moderate flux, and many more fainter. They all need to be removed to study the intrinsic properties of the diffuse emission in detail. Compact source subtraction was performed in the visibility plane for all datasets except for the ASKAP data, which was originally conducted in the image plane. We wanted to match the subtraction between MeerKAT L-band and UHF-band as closely as possible. To this end, we ensured that the compact emission of the two datasets was modelled at a similar resolution by reimaging the L-band data at \texttt{robust 0} and the UHF-band data at \texttt{robust -1.5}, which gave high-resolution images of $\sim$7$^{\prime\prime}$ for the point source models of both datasets. In addition, to ensure no extended emission was modelled, the UV-coverage in both bands was limited to the range $\ge$8.4 k$\lambda$, which corresponded to the expected restoring beam of the residuals, i.e., 30$^{\prime\prime}$. In other words, only structures smaller than 30$^{\prime\prime}$ were imaged in both bands. We experimented with different UV-cuts, and all resulted in consistent distributions of the final diffuse emission. The same imaging philosophy of self-calibration was implemented, i.e., derive a deep manual mask from an automasked initial imaging run and then restart the imaging with the manual mask. In this way, only the masked emission was subtracted. We used a common mask to ensure subtraction was consistent between bands. Figure \ref{fig:radio_optical_overlay} shows the compact and diffuse radio emission as magenta and cyan contours, respectively, all overlaid on an optical image. All magenta contours within the diffuse region were subtracted and are listed in Table \ref{tab:subtracted_srcs}. After subtraction, the residuals were reimaged at \texttt{robust 0} with three subbands over the UV-range = [0.2-18.3] k$\lambda$ with various visibility Gaussian tapers to reach a resolution close to 30$^{\prime\prime}$. Image-plane convolution was performed to reach an exact 30$^{\prime\prime}$ beam size. The noise at this resolution, estimated as the root mean square of the local residual map, was measured to be 32 $\mu$Jy/beam and 17 $\mu$Jy/beam for UHF- and L-band, respectively.

A negative hole just west of the BCG is seen in the MeerKAT L-band image, as can be seen in Figure \ref{fig:radio_optical_overlay}. This hole is also seen in \citet{2022A&A...660A..81V} and is due to the over-subtraction of a strong embedded source, J\,1327--3129a, which has residual calibration artefacts - indeed, all compact sources with residual artefacts show imperfect subtraction. Direction-dependent calibration with DDFacet and killMS \citep[following][]{2023MNRAS.520.4410T} was performed to try to eliminate these artefacts and thereby improve source subtraction. Only a marginal improvement of the artefacts was achieved, and ultimately, this negative hole persisted after subtraction. We masked this region out of all analyses presented in the following sections.

\begin{figure*}
\centering
\includegraphics[width=\textwidth]{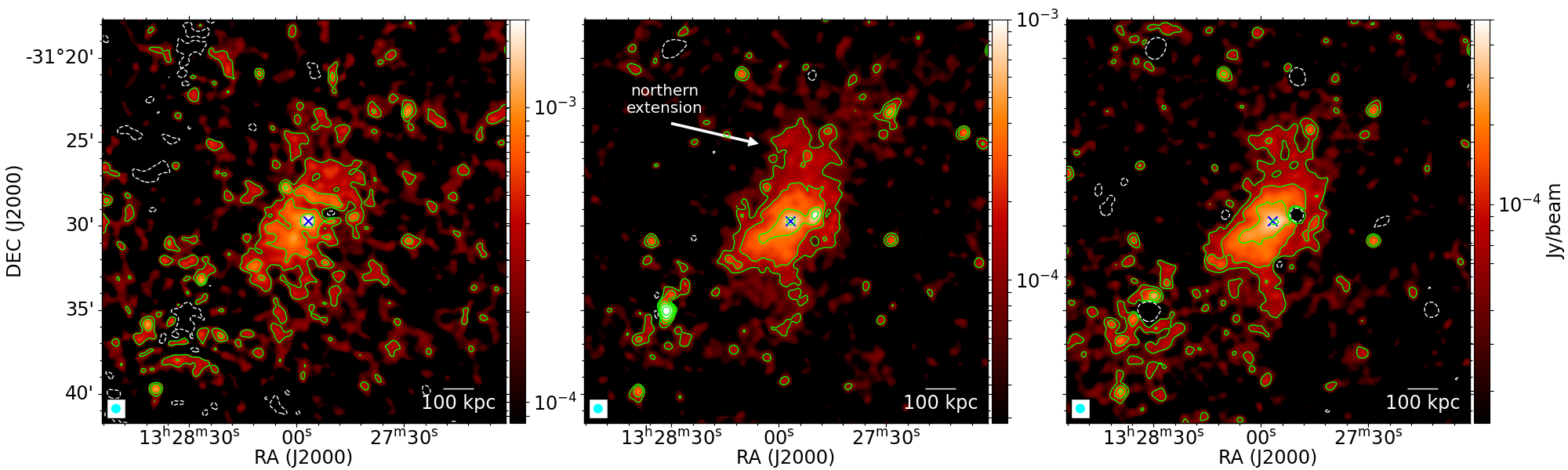}
\caption{Point source subtracted MFS continuum images of the A\,3558 diffuse emission. \textit{Left:} uGMRT 400 MHz. \textit{Middle:} UHF-band 816 MHz. \textit{Right:} L-band 1283 MHz. All images are restored to a 30$^{\prime\prime}$ beam size, with local noises of 95$\mu$Jy/beam, 32$\mu$Jy/beam, and 17$\mu$Jy/beam, respectively. The BCG is located by a blue cross. The green contours start at 3$\sigma$ and increase by a factor of two. Dashed white contours show the -3$\sigma$ level. The northern extension is labelled. A negative hole is seen in the \textit{Left} and \textit{Right} images due to the over-subtraction of the first source listed in Table \ref{tab:subtracted_srcs}.}
\label{fig:continuum_images_radio}
\end{figure*}

The uGMRT subtraction was performed similarly to MeerKAT L-band, i.e., at \texttt{robust 0} and an inner UV-cut of $\ge$8.4 k$\lambda$. Its own high-resolution mask was made, and after subtraction, the residuals were reimaged at \texttt{robust 0}, but over the full UV-range to make the detection comparable to that of MeerKAT. We found that using a similar UV-range as the MeerKAT data after subtraction (0.2-18.3 k$\lambda$) reduced the significance of the detection and reduced the recovered flux from the diffuse emission by approximately 40\%. Thus, we did not use a UV-cut when imaging the uGMRT residuals. Again, we used a combination of visibility- and image-plane Gaussian tapers to reach a final resolution of 30$^{\prime\prime}$. The noise at this resolution is 95$\mu$Jy/beam.
The final point source subtracted multi-frequency-synthesis (MFS) continuum images are shown in Figure \ref{fig:continuum_images_radio}.

\begin{figure*}
\centering
\includegraphics[width=\textwidth]{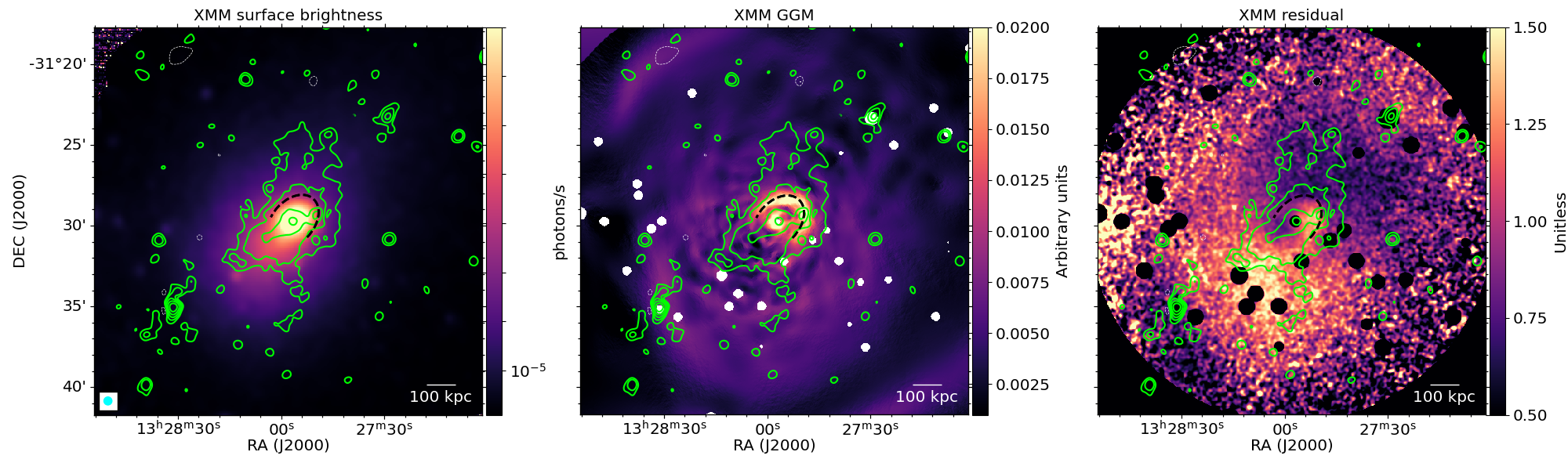}
\includegraphics[width=\textwidth]{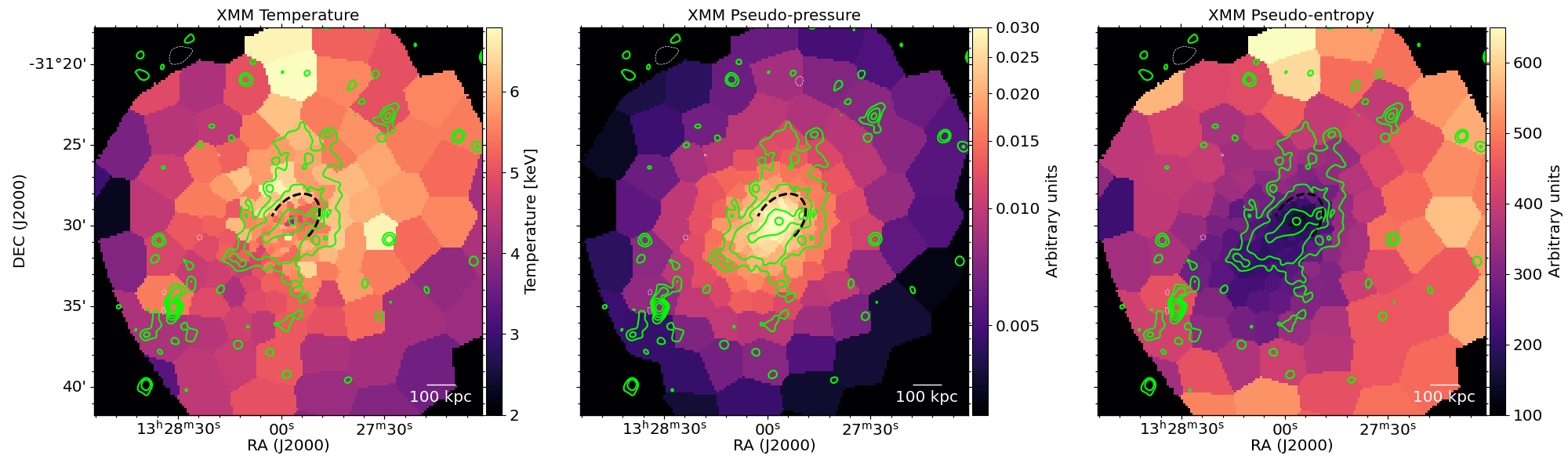}
\caption{X-ray maps with UHF-band contours overlaid (same as \textit{Figure \ref{fig:continuum_images_radio} middle}) and the cold front outlined. \textit{Top left:} surface brightness. \textit{Top middle:} GGM - white circles are masked point sources. \textit{Top right:} fractional residual brightness - black circles are masked point sources. \textit{Bottom left:} temperature. \textit{Bottom middle:} pseudo-pressure. \textit{Bottom right:} pseudo-entropy.}
\label{fig:continuum_images_xray}
\end{figure*}

\subsection{\textit{XMM-Newton}}
We analysed the archival \textit{XMM-Newton} observations of A\,3558, focusing mainly on the central pointing of Obs. ID 0107260101, which was observed on 21 January 2002 for 45 ks, using the Scientific Analysis System (SAS) version $19.1$. For the data reduction and imaging analysis, we followed the detailed procedures in \citet{2019A&A...621A..41G}. Briefly, we reprocessed the observations with the \textit{emchain} and \textit{epchain} tools to ensure the latest calibration available and performed standard pattern and flag filtering. We removed time intervals affected by flares with the \textit{mos-filter} and \textit{pn-filter} tools, resulting in a clean exposure of 43 ks (MOS1), 42 ks (MOS2), and 37 ks (pn). For each detector, we extracted source images, exposure maps and particle background images in the energy range $[0.7-1.2]$ keV, which is shown to maximise the source/background ratio, using the ESAS \textit{mos-spectra} and \textit{pn-spectra} ESAS tools \citep[see][for a description of the background model used in these tools]{2008A&A...478..615S}. We combined images of individual detectors into EPIC images and exposure maps which are used in our analysis (Figure \ref{fig:continuum_images_xray}). We detected and masked out point sources as described in \citet{2019A&A...621A..41G}. \\
Concerning the thermodynamic maps shown in the bottom panels of Figure \ref{fig:continuum_images_xray}, we relied on the results presented in \citet{2007A&A...463..839R}. Briefly, we applied an adaptive binning algorithm based on Voronoi tessellation \citep{2003MNRAS.342..345C} to ensure a constant signal-to-noise ratio throughout the images. In each bin, we fit the count-rates in five energy bands to derive the temperature $T$ and normalisation $K$ and combine them to derive the \textit{pseudo-pressure} and \textit{pseudo-entropy} quantities as:
\begin{equation}
P \equiv T\left( \frac{K}{N_p}\right)^{1/2} \\
s \equiv T\left( \frac{K}{N_p}\right)^{-1/3},
\label{eqn:1}
\end{equation}
where $N_p$ is the number of pixels in each bin.

\section{Continuum images and analysis}
\label{sec:continuum_imgs}

\subsection{Radio}

Figure \ref{fig:continuum_images_radio} shows the point source subtracted radio surface brightness continuum maps used in this work. They are convolved to a 30$^{\prime\prime}$ beam size to recover as much of the diffuse emission as possible and are centred on the location of the BCG. The diffuse emission of interest is detected in all datasets - having a morphology similar to that shown in \citet{2022A&A...660A..81V}, i.e., elliptical and generally following the X-ray surface brightness distribution, which itself is shown in the top-left panel of Figure \ref{fig:continuum_images_xray} and described in the next section. A faint extension to the north is recovered in the MeerKAT data, and just barely seen in the uGMRT counterpart, which was not seen in \citet{2022A&A...660A..81V}. This extension increases the projected largest linear size (LLS) to $\sim 550$ kpc.  

\begin{figure*}
\centering
\includegraphics[width=0.38\textwidth]{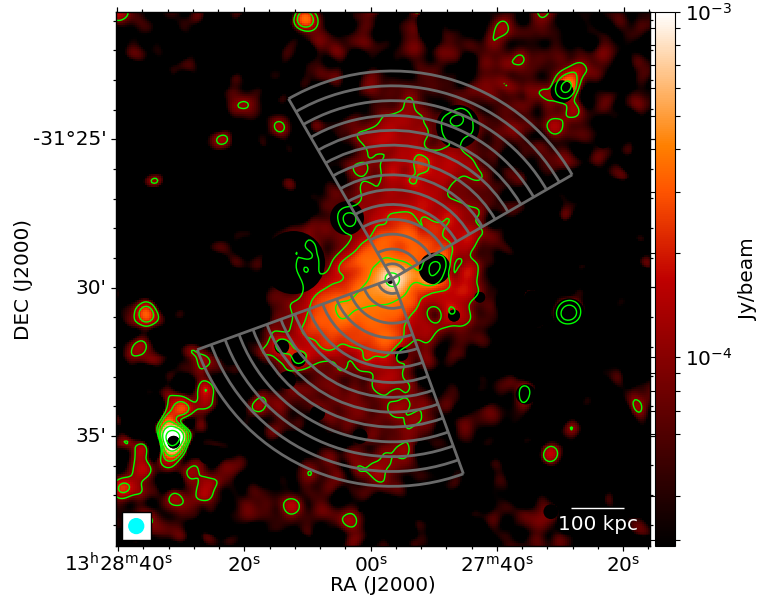}
\includegraphics[width=0.30\textwidth]{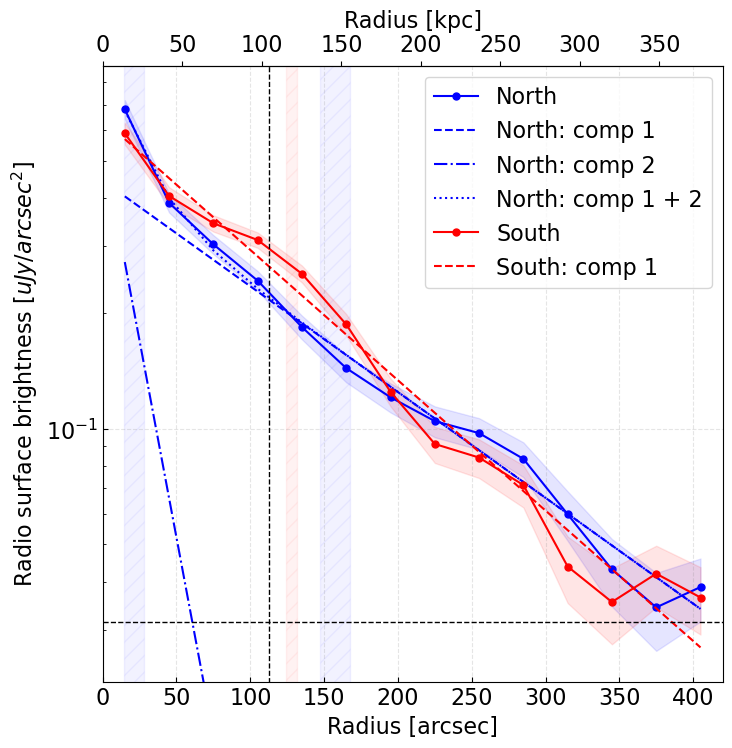}
\includegraphics[width=0.30\textwidth]{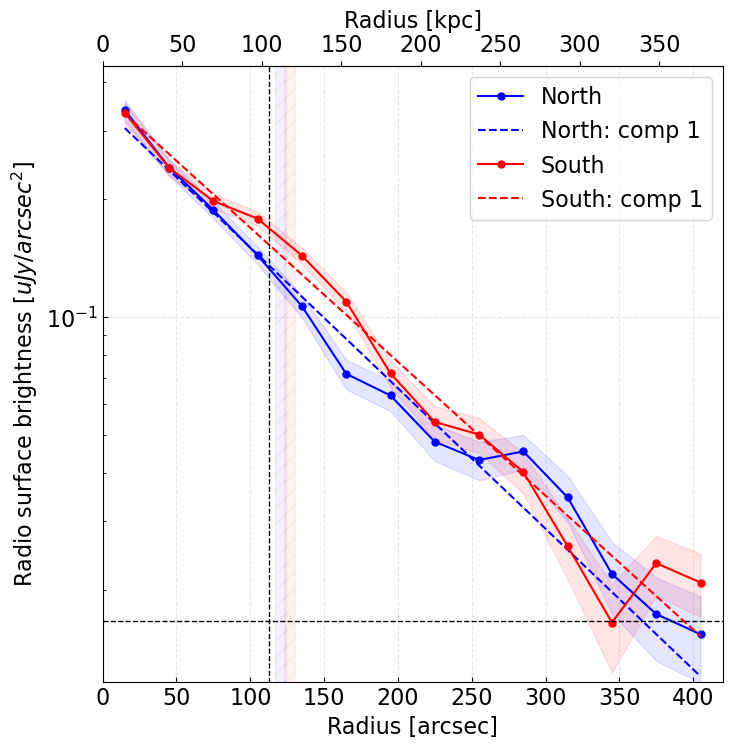}
\caption{A\,3558 UHF- and L-band surface brightness profiles. \textit{Left:} Same as Figure \ref{fig:continuum_images_radio} (\textit{Middle}) but with sectors overlaid showing the sampling of the radial profiles in the north and south directions. Some obvious residual compact sources are masked out. \textit{Middle:} UHF-band profiles. \textit{Right:} L-band profiles. The measurements in the northern sector are coloured red and the southern sector blue. Shaded regions show the 1$\sigma$ uncertainties on the surface brightness measurements. The dashed (and dash-dotted) lines show the best-fit exponential profiles for each component, and the vertical shaded regions show the best-fit e-folding radii. The vertical dashed line shows the cold front position from \citet{2007A&A...463..839R}. The horizontal dashed lines show the noise levels. } 
\label{fig:radial_profiles_sb}
\end{figure*}

We measured the radio flux density within the 3$\sigma$ contour at each frequency and calculate the uncertainty as
\begin{equation}
\label{eqn:01}
\Delta S = \sqrt{ \left( \sigma_{\rm cal} S \right)^{2} + \left( \sigma \sqrt{N_{\rm beam}} \right)^{2} + \sigma_{\rm sub}^{2} }.
\end{equation}
where $\sigma_{\rm cal}$ is the percentage calibration uncertainty, $S$ is the measured radio flux density, $\sigma$ is the local noise given in the caption of Figure \ref{fig:continuum_images_radio}, $N_{\rm beam}$ is the number of beams contained within the 3$\sigma$ region and, 
\begin{equation}
\label{2}
\sigma_{\rm sub}^{2} = \sum_{\mathclap{s=1}}^{N} \left(I_{\rm s} \times N_{\rm beam,\,s}\right)^{2},
\end{equation}
is the subtraction uncertainty. Here, $I_{\rm s}$ is the mean residual surface brightness of the diffuse emission within the $s^{\rm th}$ source region, and $N_{\rm beam,\,s}$ is the number of beams within that region. The flux density for the uGMRT and ASKAP MFS images and three equal subbands in each MeerKAT band are given in Table \ref{tab:flux_measurements}. The subbands are chosen to highlight the overlap between the UHF- and L-bands at 1 GHz. The ASKAP and L-band flux densities used here are consistent with those presented in \citet{2022A&A...660A..81V}.  

\begin{table}
\caption{Flux density measurements of the A3558 diffuse emission for this work.}
\centering
\begin{tabular}{ccc}
\hline
$\nu$ (MHz) & Array & Flux density (mJy)\\
\hline
400 & uGMRT (Band-3)& 54.85$\pm$5.87\\
635 & MeerKAT (UHF-band) & 30.35$\pm$3.38\\
816 & MeerKAT (UHF-band)  & 25.92$\pm$2.99\\
887 & ASKAP & 26.05$\pm$5.21\\
997 & MeerKAT (UHF-band)  & 19.07$\pm$2.42\\
998 & MeerKAT (L-band) & 17.45$\pm$1.93\\
1283 & MeerKAT (L-band) & 13.24$\pm$1.56\\
1569 & MeerKAT (L-band) & 11.32$\pm$1.40\\
\hline
\end{tabular}
\label{tab:flux_measurements}
\end{table}

\subsection{X-ray}

The images of the thermal quantities of A\,3558 as derived from XMM EPIC are shown in Figure \ref{fig:continuum_images_xray}. The thermal plasma in A\,3558 has been shown to have three cold fronts indicating gas sloshing throughout the cluster volume \citep{2007A&A...463..839R,2023MNRAS.526L.124M}. In particular, the innermost cold front, at 113$^{\prime\prime}$ (105 kpc) northwest of the central density peak, lies within the diffuse radio emission. To highlight this innermost cold front and its spatial coincidence with the diffuse radio emission, we show in the middle-top panel of Figure \ref{fig:continuum_images_xray} the adaptive Gaussian Gradient Method \citep[GGM,][]{2022A&A...661A..36S} image of the X-ray surface brightness. The intensity of this image traces the spatial gradient in surface brightness and is used to detect edge features such as cold fronts or shocks. The \textit{adaptive} implementation of this method uses a multiscale approach to ensure a minimum signal-to-noise ratio (we used the default setting of S/N = 32) in each smoothing kernel, resulting in sharp, fine details being preserved in bright regions and smoother, noise-suppressed gradients in fainter regions. We find that the peak of this GGM map indeed coincides with the location of the innermost, northwest, cold front. 

To highlight the sloshing features often associated with cold fronts, we used the \textit{PyProffit} package \citep{2020OJAp....3E..12E} to model the thermal plasma as an elliptical double-$\beta$ distribution. The observed surface brightness was then divided by the best-fit model to extract the underlying sloshing features. The image showing this ratio (top right panel of Figure \ref{fig:continuum_images_xray}) traces local under/overdensities of the thermal ICM and reveals the classic spiral signatures of gas sloshing. The best-fit parameters are given in Table \ref{tab:double_beta_mod}, including the single $\beta$ value used for both components, the two core radii in arcminutes, the ratio between the two normalisations, the logarithm of the normalisation of the first model, and the logarithm of the X-ray background. We note that the inner beta core radius is just larger than the radius of the northern cold front stated in \citet{2007A&A...463..839R}, showing that this feature does indeed trace the cold front. The bottom row of Figure \ref{fig:continuum_images_xray} also shows the X-ray temperature, pseudo-pressure and pseudo-entropy maps from \citet{2007A&A...463..839R}. The radio contours are overlaid onto these X-ray maps and suggest a clear connection between the thermal and non-thermal emissions. For example, the orientation of the radio and X-ray surface brightness and thermodynamic quantities are similar, suggesting that they are both manifestations of ICM processes. Furthermore, the brightest radio emission is confined within the cold front, with a low surface brightness extension to the north in a hot, X-ray deficient region of high pressure and high entropy. The analysis and interpretation of such features and the connection between the thermal and non-thermal emission are presented and discussed in the following sections.

\begin{table}
\caption{Best-fit parameters for radio surface brightness radial profiles. I$_{0}$ has units of $\mu$Jy/arcsec$^{2}$ and r$_{e}$ of arcseconds.}
\centering
\begin{tabular}{c|cccc}
\hline
& \multicolumn{2}{c}{UHF-band} |& \multicolumn{2}{c}{L-band} \\
\hline
\multirow{2}{*}{North Comp 1} & I$_{0}$ & r$_{e}$ & I$_{0}$ & r$_{e}$ \\
& 0.44$\pm$0.04 & 158$\pm$10 & 0.34$\pm$0.01 & 121$\pm$4 \\
\hline
\multirow{2}{*}{North Comp 2} & I$_{0}$ & r$_{e}$ & - & - \\
& 0.55$\pm$0.14 & 21$\pm$7 & - & - \\
\hline
\multirow{2}{*}{South Comp 1} & I$_{0}$ & r$_{e}$ & I$_{0}$ & r$_{e}$ \\
& 0.64$\pm$0.02 & 129$\pm$4 & 0.37$\pm$0.01 & 127$\pm$4\\
\hline
\end{tabular}
\label{tab:radio_radial_profiles}
\end{table}


\begin{table}
\caption{Best-fit parameters of the X-ray \textit{PyProffit} Double Beta model, parameterised by a single $\beta$ value, the two core radii (arcminutes), the ratio between the two components, the logarithm of the first normalisation and the logarithm of the X-ray background.}
\centering
\begin{tabular}{ccc}
\hline
$\beta$ & $r_{c1}$ & $r_{c2}$ \\
1.08$\pm$0.08 & 2.13$\pm$0.13 & 9.1$\pm$0.5\\
\hline
ratio & norm & bkg\\
0.40$\pm$0.01 & -1.33$\pm$0.01 & -3.06$\pm$0.05\\
\hline
\end{tabular}
\label{tab:double_beta_mod}
\end{table}

\begin{figure}
\centering
\includegraphics[width=\columnwidth]{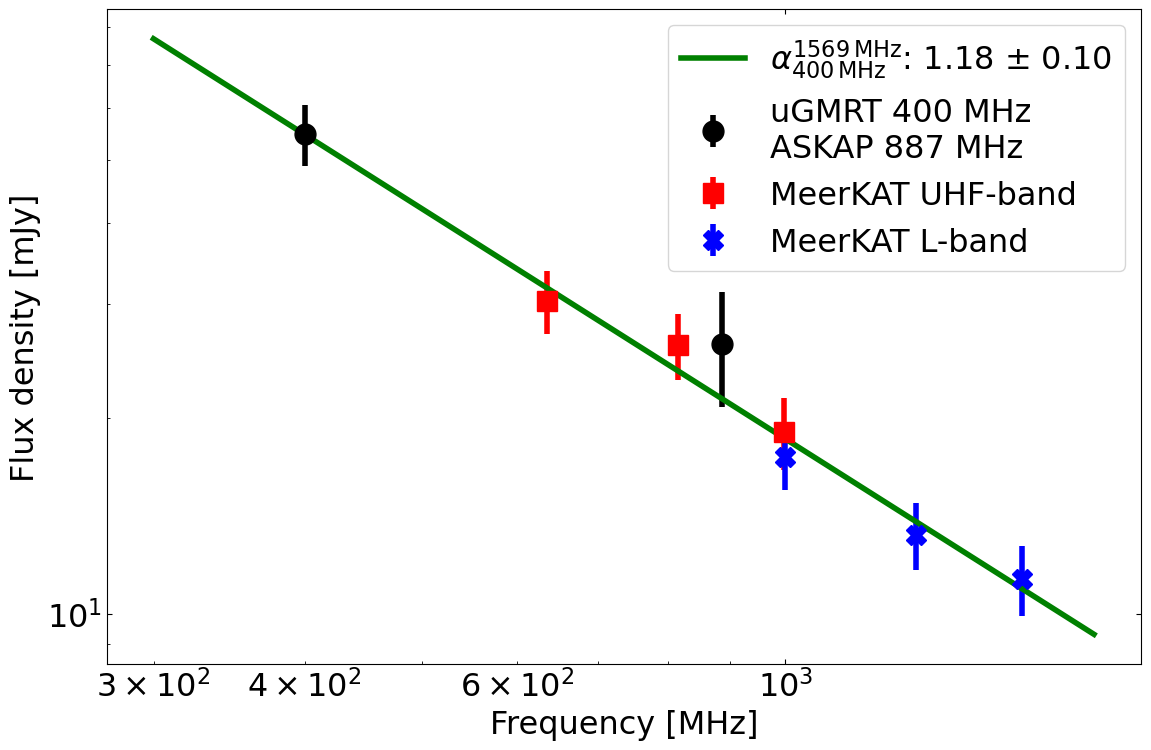}
\caption{A3558 integrated spectrum. Red squares and blue crosses show the MeerKAT UHF- and L-band data points, respectively. The black circles show uGMRT at 400 MHz and ASKAP at 887 MHz. }
\label{fig:int_spec}
\end{figure}

\begin{figure*}
\centering
\includegraphics[width=0.62\textwidth]{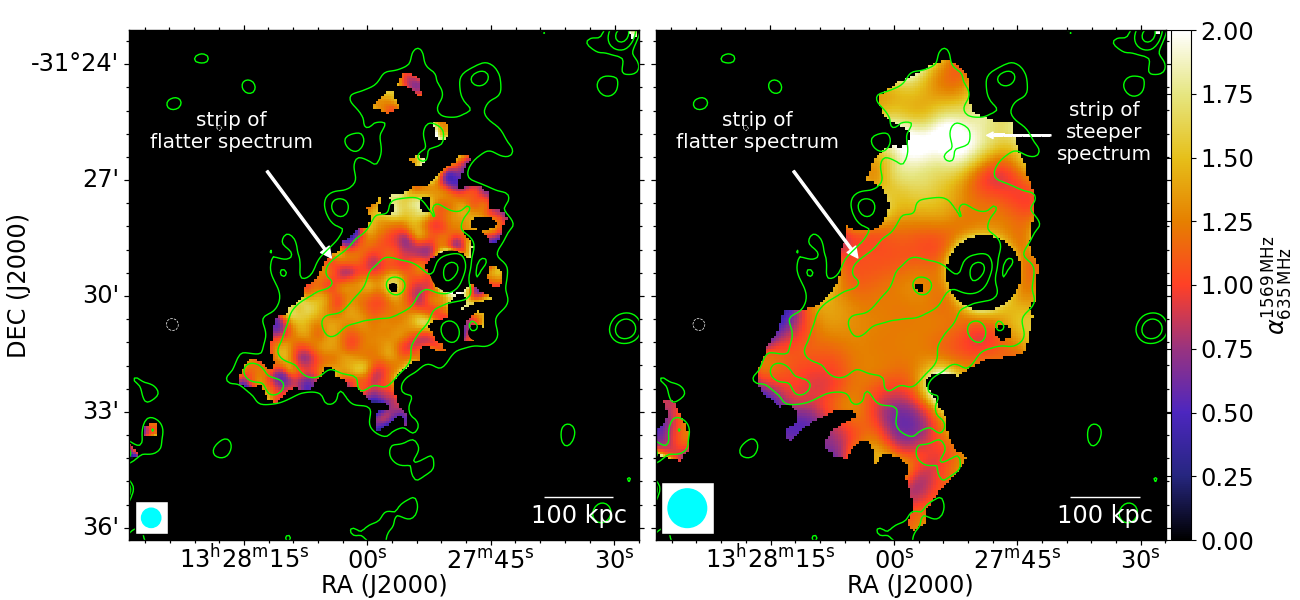}
\includegraphics[width=0.365\textwidth]{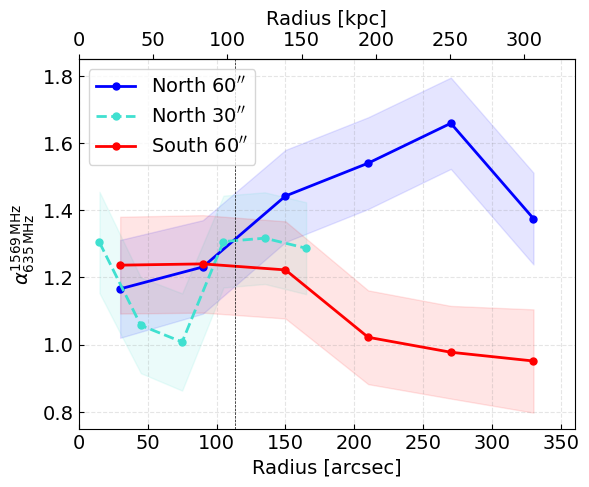}
\caption{A3558 spectral index map and radial profile. \textit{Left:} spectral index map at 30$^{\prime\prime}$ resolution with UHF-band contours overlaid. \textit{Middle:} same as \textit{left} but at 60$^{\prime\prime}$. \textit{Right:} radial profile similar to Figure \ref{fig:radial_profiles_sb} for the 60$^{\prime\prime}$ resolution spectral index map where blue is the northern sector and red is the southern sector. Light blue shows the profile for the northern sector in the 30$^{\prime\prime}$ resolution spectral index map and highlights the strip of flatter spectrum.}
\label{fig:spec_index_map}
\end{figure*}

\begin{figure*}
\centering
\includegraphics[width=0.65\textwidth]{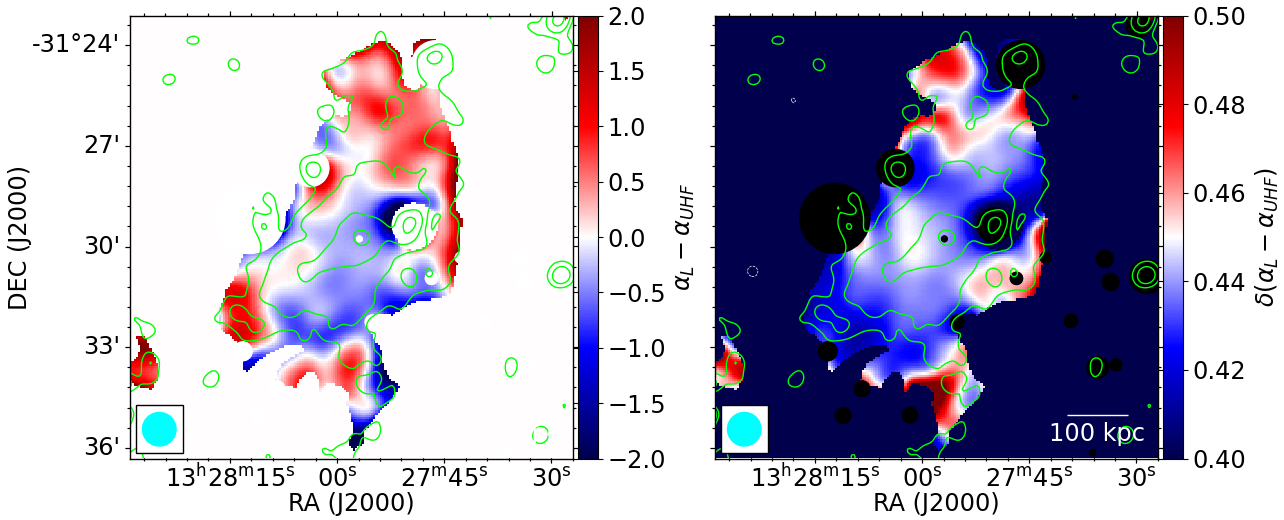}
\includegraphics[width=0.34\textwidth]{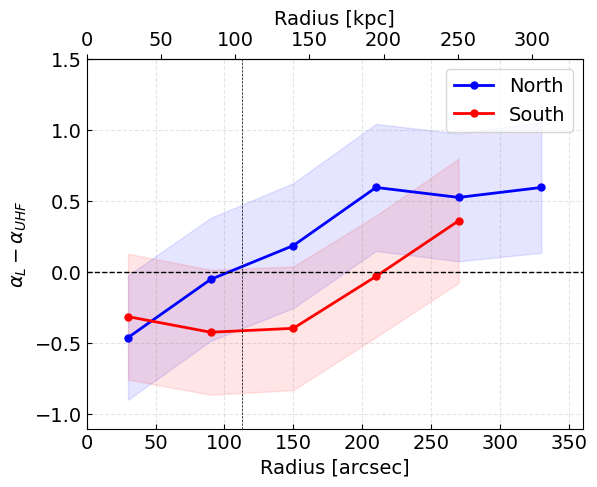}
\caption{A\,3558 spectral curvature map and radial profile at 60$^{\prime\prime}$. \textit{Left:} curvature map. \textit{Middle:} uncertainty map. \textit{Right:} radial profile.}
\label{fig:curv_index_map}
\end{figure*}

\section{Analysis of the diffuse radio emission}
\label{sec:analysis}

In this section, we present a detailed analysis of the radio properties of the diffuse emission in A\,3558, including radial profiles and spectral properties. 
To this aim we use our highest fidelity MeerKAT UHF- and L-band point source subtracted images (Figure \ref{fig:continuum_images_radio}'s \textit{Middle} and \textit{Right} panels) to make radial profiles of the diffuse emission, and the subband images of these datasets for an integrated spectrum, spectral index and curvature maps of the diffuse emission.

\subsection{Radio radial profiles}
\label{sec:radial_profiles}


Figure \ref{fig:radial_profiles_sb} shows the radial profiles for the MeerKAT UHF- and L-band diffuse emission in two conic sections along the orientation axis. To properly cover the northern and southern extension of the diffuse emission, the northern section is drawn between 30-120$^{\circ}$ from the horizontal, and the southern section from 200-290$^{\circ}$. The annuli all have a radius equal to the size of the restored beam (30$^{\prime\prime}$), making the measurements independent. For simplicity, we mask out the main contaminating point sources, which may still have some residual compact emission. Then, the surface brightness and uncertainties are calculated similarly to the flux densities, but each is divided by the annulus area. At first glance, the shapes of the profiles are fairly consistent across direction and frequency, but a few differences are hinted at. Specifically, in the north in the UHF-band, a slight increase in brightness is seen in the first annulus. Thereafter, the southern profile experiences a slightly higher brightness in the radius range 75$^{\prime\prime}$-200$^{\prime\prime}$ (70-185 kpc). Subsequently, the northern extension is seen as a surplus of emission compared to the southern profile at $\ge$250$^{\prime\prime}$ ($\ge$230 kpc) in both bands. 


To formally characterise the profiles, we fit a single exponential in the conic sections following \citet{2009A&A...499..679M} of the form 
\begin{equation}
\label{3}
I_{R}(r) = I_{0}\exp{(\,-\frac{r}{r_{e}})\,},
\end{equation}
where $I_{0}$ is the peak surface brightness and $r_{e}$ the e-folding radius. We use the \textit{scipy.optimize.curve\_fit} function to perform the fit, and Table \ref{tab:radio_radial_profiles} gives the best-fit parameters. These parameters depend slightly on the choice of sample profile but generally agree within the stated uncertainties. All profiles can be characterised by a single exponential within the observational uncertainties, except for that in the UHF-band in the north. Here, the surplus of emission mentioned above allows us to constrain a two-component exponential, where the second component characterises the core region. Local deviations from the best-fit may be real enhancements/decrements of the diffuse emission or attributed to residual contamination from embedded compact sources. We cannot constrain whether the northern extension constitutes a second (or third) component due to the sensitivity limits of our images. Indeed, it may just be the brightest part of a larger, more coherent component. Deep low-frequency observations are required to constrain whether multiple components are present at large radii. 

\subsection{Spectral Properties}

The spectral index, $\alpha$, is key for probing the mechanisms responsible for the origin of the diffuse emission. By analysing the spectral index, one can estimate the energy distribution and particle (re)-acceleration mechanisms of the cosmic ray electrons (CRe) responsible for the non-thermal emission, and the average magnetic field strength of the ICM \citep{2014IJMPD..2330007B}. For example, steepening of the spectral index at higher frequencies in the form of a break in the energy spectrum suggests radiative cooling dominates over (re)-acceleration at high energies, typical of turbulent (re)-acceleration models \citep{2014IJMPD..2330007B}. With high-quality radio data from $\sim$400 MHz to $\sim$1500 MHz, we characterise the spectral nature of A\,3558 over a $\sim$1 GHz frequency range.

\subsubsection{Integrated spectrum}


The flux densities of the diffuse emission at the uGMRT and ASKAP central frequencies and three subbands in each of the MeerKAT bands are given in Table \ref{tab:flux_measurements}. We fitted a single power-law to the spectrum using the \textit{scipy.optimize.curve\_fit} function and took the uncertainties as the square root of the diagonal terms of the covariance matrix. The resulting integrated spectrum is shown in Figure \ref{fig:int_spec}. We find an integrated spectral index of $\alpha_{\rm 400\,MHz}^{\rm 1500\,MHz} = 1.18 \pm 0.10$. This is significantly flatter than the ultra-steep-spectrum (USS) given in \citet{2022A&A...660A..81V} of $\alpha_{\rm 887\,MHz}^{\rm 1283\,MHz} = 2.3 \pm 0.4$. However, such a spectral index is reproduced in this work if only the ASKAP and MeerKAT data are fitted. Hence, we suggest that the USS is caused by the ASKAP measurements being biased to higher flux densities by inaccurate point source subtraction. With our spectral index, we determine a radio power of $P_{\rm 1.4\,GHz} = 6.8\pm0.9 \times 10^{22}$ W/Hz. 

We also measure the BCG flux density over a similar frequency range as above. These are listed in Table \ref{tab:bcg_flux_measurements}, resulting in a spectral index of $\alpha_{\rm 400\,MHz}^{\rm 1658\,MHz} = 1.04\pm0.06$, shown in Figure \ref{fig:bcg_int_spec}, and radio power $P_{\rm 1.4\,GHz} = 2.57\pm0.26 \times 10^{22}$ W/Hz, consistent with \citet{2018A&A...620A..25D}


\subsubsection{Spectral index map}
\label{sec:spx}

The resolved spectral index map between the six MeerKAT subbands at 30$^{\prime\prime}$ and 60$^{\prime\prime}$ resolution are shown in Figure \ref{fig:spec_index_map}, with the associated uncertainty maps in Figure \ref{fig:spec_index_uncert}. The MeerKAT bands have consistent UV ranges, i.e., same UV cuts and comparable coverage of the UV plane, and point source subtraction, leading to the highest fidelity spectral index map. The map is made by fitting the same power-law as above, again with \textit{scipy.optimize.curve\_fit} function, but now on a pixel-by-pixel basis. This results in a spectral distribution that is generally consistent with the integrated spectrum but with several defining spatial features. In the 30$^{\prime\prime}$ map, a clear spiral-shaped strip of flatter spectrum, a few beams in length and one beam across, is seen just north of the BCG. Unfortunately, a contaminating point source (J1327-3129a, see Figure \ref{fig:radio_optical_overlay}) is positioned at this location, obscuring the westernmost part of the strip. Further out, in the 60$^{\prime\prime}$ map, a strip of steeper spectrum is seen. We interpret these features in the context of gas sloshing and particle (re)-acceleration to connect the energetics of the non-thermal ICM to the dynamics of the thermal plasma and discuss the origin of the radio diffuse emission in Section \ref{sec:gas_sloshing}. 

The right panel of Figure \ref{fig:spec_index_map} shows radial profiles along the orientation of the diffuse emission using similar annuli to Figure \ref{fig:radial_profiles_sb}. The annuli are again one beam in radius corresponding to the 30$^{\prime\prime}$ and 60$^{\prime\prime}$ maps, respectively. On large scales, in the 60$^{\prime\prime}$ map, the northern region exhibits significant radial steepening, starting from $\alpha \sim 1.2$ inside the cold front, and increasing consistently until a peak at $\sim$275$^{\prime\prime}$ (255 kpc) with $\alpha \sim 1.7$ and dips thereafter to $\alpha \sim 1.4$. In the southern part of the cluster, the spectrum stays fairly constant at $\alpha \sim 1.2$ until $\sim$150$^{\prime\prime}$ (140 kpc) and then gradually flattens thereafter to $\alpha \sim 1.0$. We also plot the northern profile from the 30$^{\prime\prime}$ map, which highlights the strip of flat spectrum on small scales inside the cold front. Finally, the 30$^{\prime\prime}$ profile in the south is consistent with the 60$^{\prime\prime}$ counterpart and so is not shown.

\begin{figure*}
\centering
\includegraphics[width=0.35\textwidth]{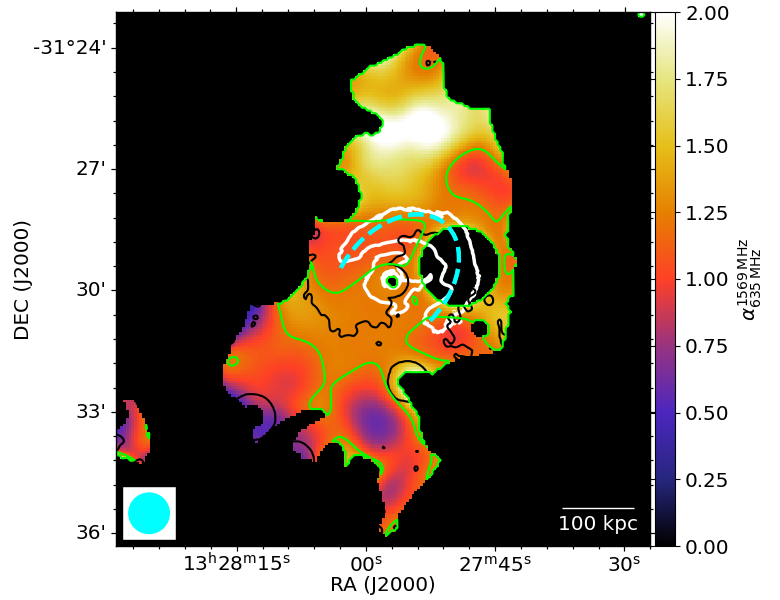}
\includegraphics[width=0.32\textwidth]{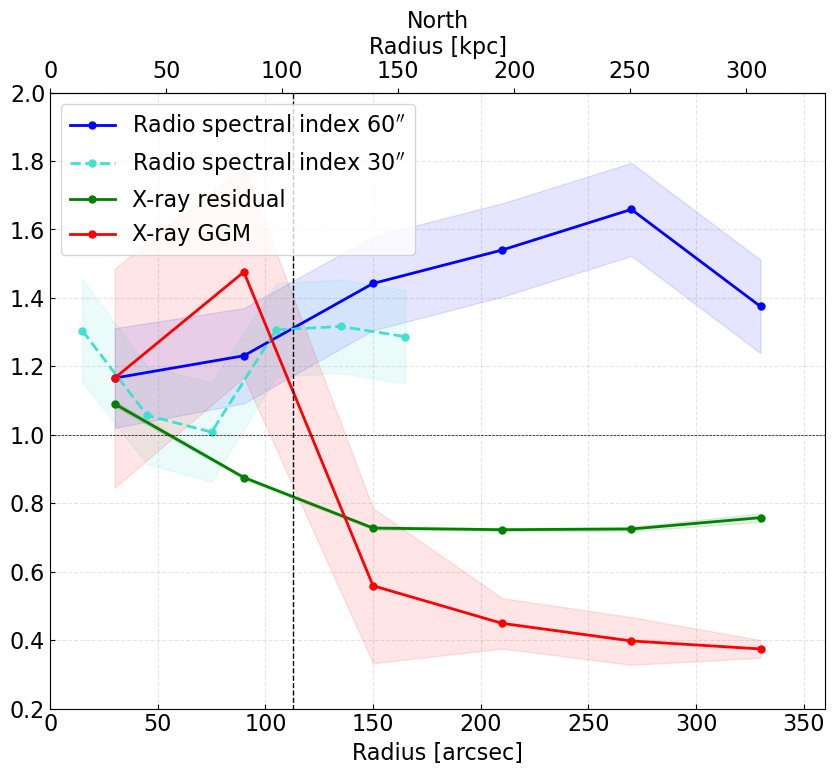}
\includegraphics[width=0.32\textwidth]{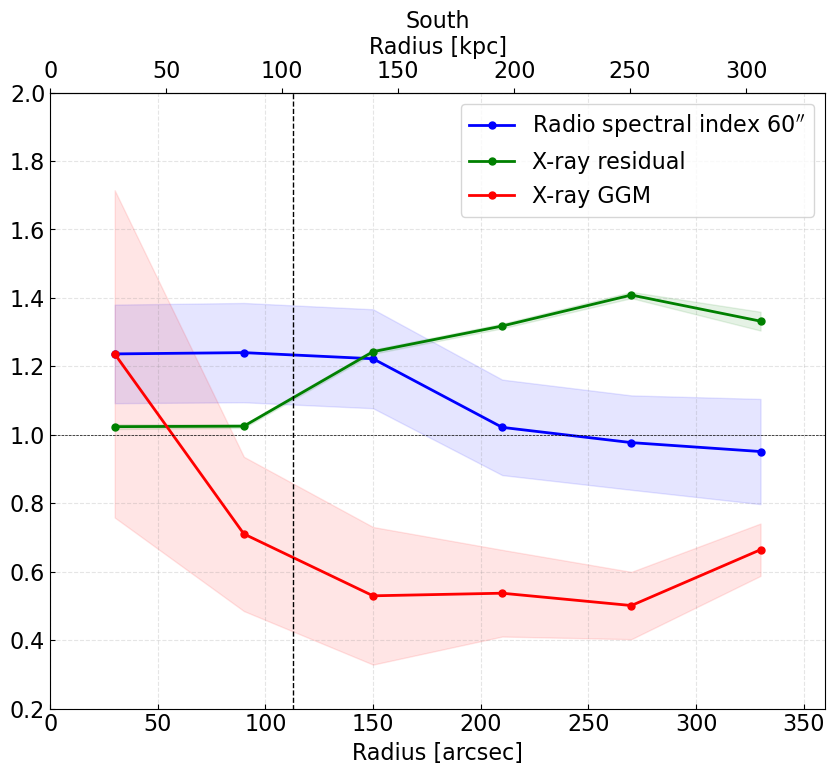}
\caption{A\,3558 spectral index profiles against X-ray GGM and residual surface brightness maps. \textit{Left}: spectral index map from Figure \ref{fig:spec_index_map} with a green contour overlaid at $\alpha=1.2$. The white contours show the X-ray GGM peak region, and the cyan dashed ellipse the position of the X-ray cold front. The black contour shows the unity level of the X-ray residual map, with regions less than unity to the south and greater than unity to the north. \textit{Middle:} plot comparing the spectral index (blue) to the X-ray GGM (red) and residual (green) profiles in the north. \textit{Right:} same as \textit{middle} but for the southern region.}
\label{fig:radial_profiles_alpha}
\end{figure*}

\subsubsection{Spectral curvature}

Spectral curvature is the change in spectral index across frequency and can be used to identify regions with different (re)-energisation processes \citep[e.g.][]{2014IJMPD..2330007B,2024ApJ...975..125R}. The spectral curvature map is produced using two independent spectral index maps, the UHF and L in-band spectral index maps, where the curvature is defined as $\alpha_{\rm L} - \alpha_{\rm UHF}$, or $\alpha_{\rm 998\,MHz}^{\rm 1569\,MHz} - \alpha_{\rm 635\,MHz}^{\rm 997\,MHz}$. Here, a positive curvature indicates spectral steepening, a classic signature of electron ageing, which is expected also in the turbulent re-acceleration scenario. Figure \ref{fig:curv_index_map} shows the curvature map and associated uncertainty maps. Since the two individual spectral maps are independent of each other, the curvature uncertainty is derived as the simple propagation of errors from the two in-band maps. The same annuli as before are used to derive the radial profile. Although the uncertainties are large, we can still discern some unique behaviour. In the south, the curvature remains fairly uniform at negative or no curvature until $\sim150^{\prime\prime}$ (140 kpc), suggesting a consistent and uniform energy injection, and then gradually increases to positive (i.e. spectral steepening). The north also starts at negative curvature but increases immediately, becoming positive at $\sim200^{\prime\prime}$ (185 kpc) and staying constant throughout the northern extension. This suggests electron ageing in the northern and southern extensions compared to the central region of the diffuse emission.

\section{Correlations between the non-thermal and thermal emission}
\label{sec:correlations}

\begin{figure}
\centering
\includegraphics[width=\columnwidth]{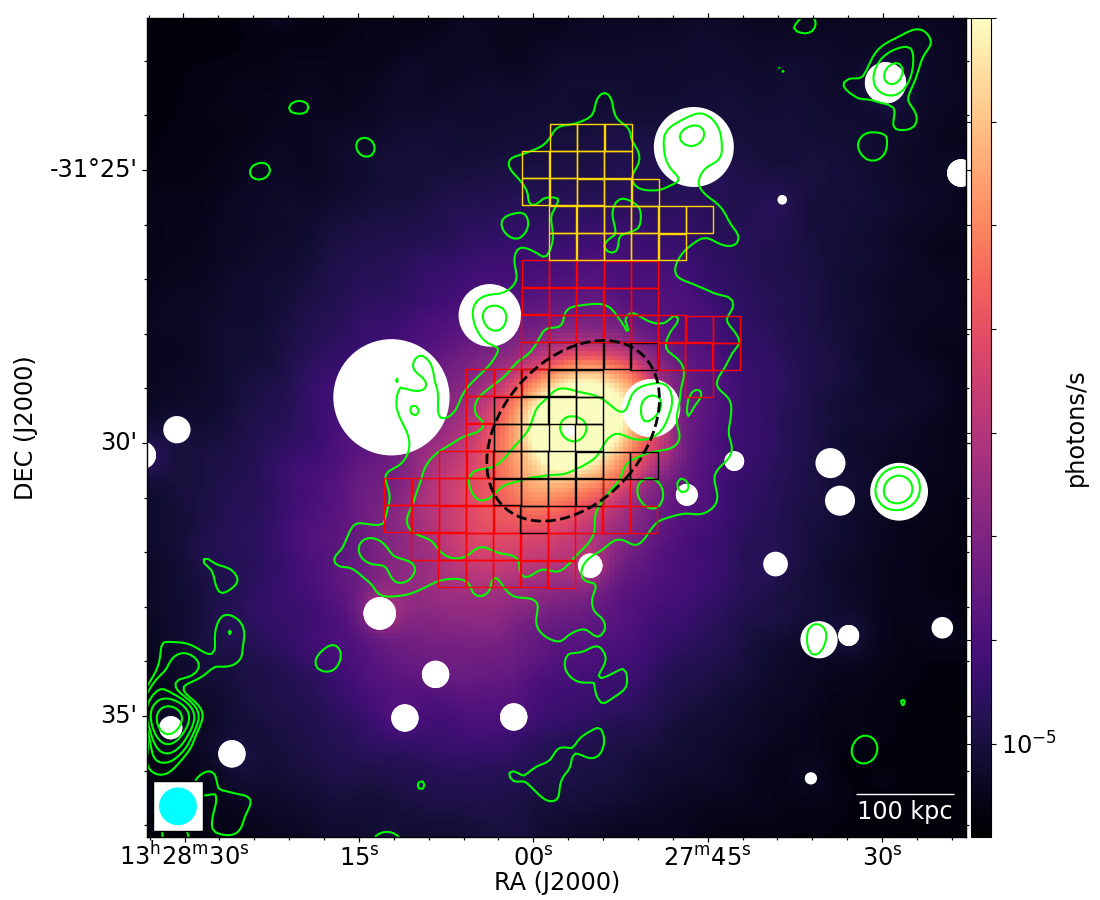}
\includegraphics[width=0.49\columnwidth]{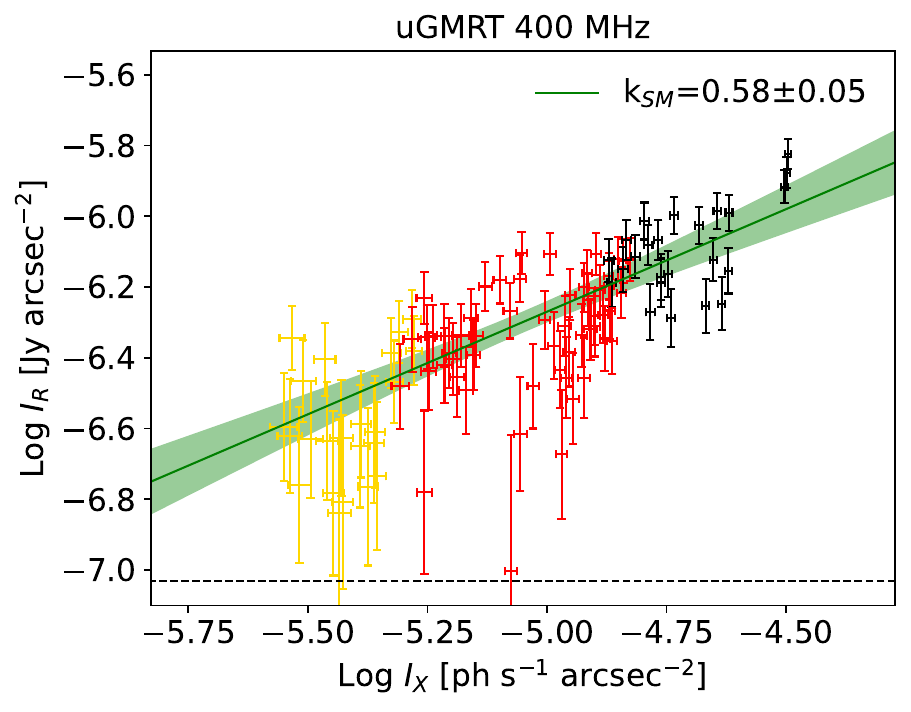}
\includegraphics[width=0.49\columnwidth]{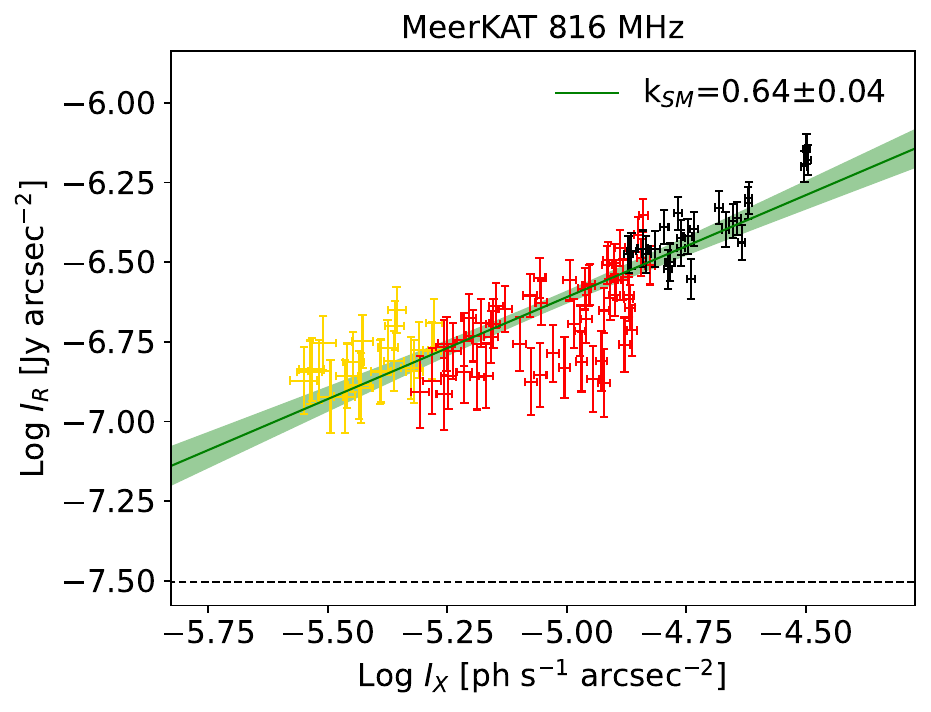}
\includegraphics[width=0.49\columnwidth]{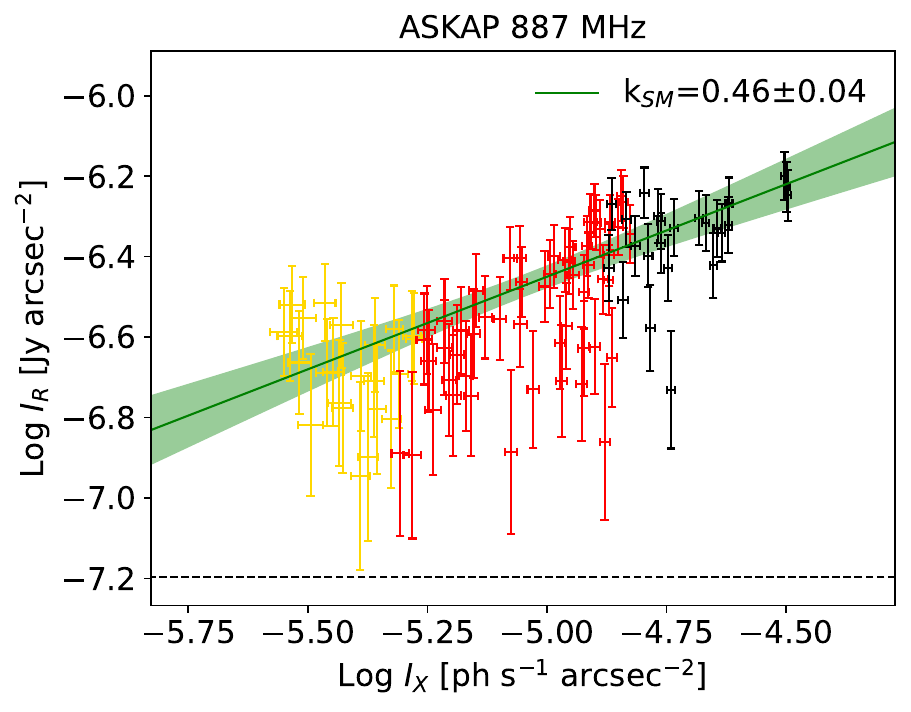}
\includegraphics[width=0.49\columnwidth]{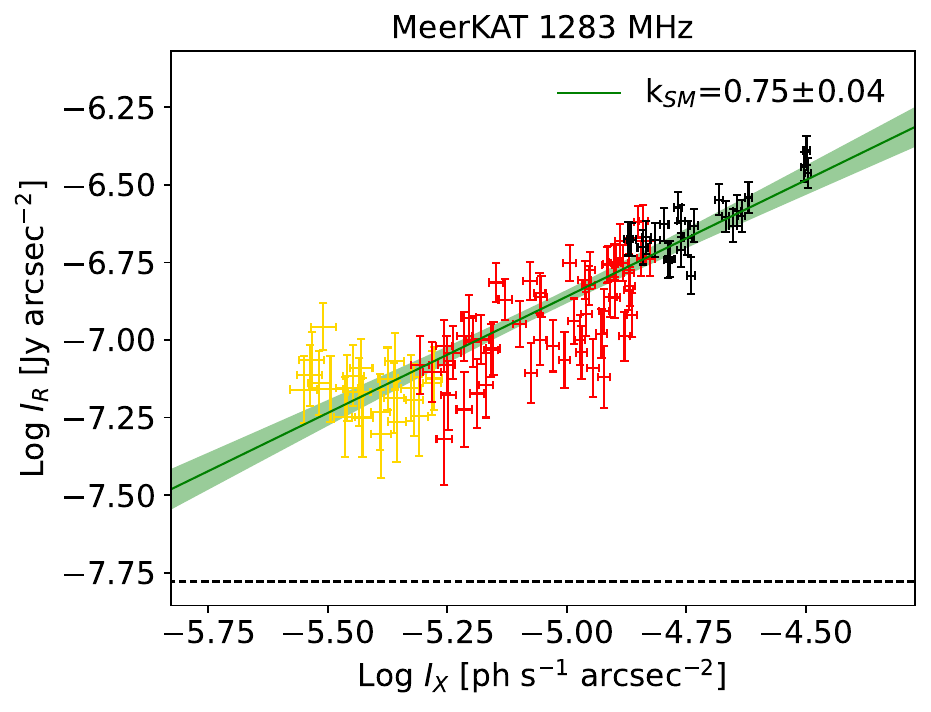}
\caption{A\,3558 $I_{R}/I_{X}$ correlation analysis. \textit{Top:} X-ray surface brightness map shown in Figure \ref{fig:continuum_images_xray} with UHF contours overlaid. The mesh grid shows the sampling distribution for the surface brightness point-to-point analyses, where black squares show the region inside the cold front, red the region outside the cold front and yellow the northern extension. The white circles are masked point sources. \textit{Middle and bottom:} correlation planes for the various data sets. The data points are colour-coded to match the cells in the colour map. The best-fit slope using all data points in each plane is given in green with the shaded region the 1$\sigma$ uncertainty. The horizontal line shows the noise level.}
\label{fig:ptp-sb}
\end{figure}

Studying the relationship between the thermal and non-thermal emissions provides valuable insight into the overall dynamics of the environment and how the energy injected during merger activities affects the different components of the ICM.

\subsection{Radial profiles} 
\label{sec:radial_profiles1}


We compare the radial profiles of the spectral index map to the X-ray GGM and residual maps to showcase trends that directly link changes in the thermal emission with the energetics of the non-thermal emission. Figure \ref{fig:radial_profiles_alpha} highlights the spatial coincidence between the spectral index and the X-ray GGM and residual maps. We show radial profiles of $\alpha$ against the X-Ray GGM and residual surface brightness in the same annuli as Figure \ref{fig:radial_profiles_sb}. The X-ray GGM profiles have been normalised to match the spectral index for ease of comparison. The region of flatter spectrum just north of the BCG mentioned in Section \ref{sec:spx} is coincident with the peak region of the X-ray GGM map. The spectrum then steepens across the cold front and peaks at $\sim$275$^{\prime\prime}$ ($\sim$255 kpc), as mentioned in Section \ref{sec:spx}, while the GGM and X-ray residual surface brightness decreases. In fact, the Pearson and Spearman correlation coefficients for the spectral index versus X-ray GGM in the north are -0.8 and -0.6, and for spectral index versus residual X-ray surface brightness are -0.8 and -0.9, respectively, clearly indicating a strong negative correlation. On one hand, these relations directly link the processes responsible for the thermal cold front and the (re)-energisation of the non-thermal component. On the other hand, they link the local under-density due to gas sloshing to the ageing of the synchrotron-emitting electrons. Regarding the southern region, the relations are not quite as obvious. Here, the radio spectrum stays uniform until the e-folding radius, while the X-ray GGM decreases and the X-ray residual increases. Thereafter, the spectrum becomes flatter while the GGM stays fairly uniform and the X-ray residual continues to increase. This suggests opposite trends between the X-ray GGM and residual quantities. Indeed, the Pearson and Spearman coefficients for the spectral index versus X-ray residual are both -0.9, and spectral index versus GGM are both 0.5 (with p-values of 0.3). The latter reiterates the connection between the local over-density of gas sloshing to (re)-energisation, while the exceptional trend in the south against the X-ray GGM and its positive correlation with large p-value highlights the complexity of the region and is interpreted as the (re)-energisation caused by gas sloshing dominating over ageing in this direction thereby flattening the radio spectrum instead of steepening it.

\subsection{Point-to-point analysis: Surface brightness}

It is a common tool to study the thermal/non-thermal interplay in galaxy clusters with diffuse emissions via point-to-point analyses - for example, see \citet{2020A&A...640A..37I,2021A&A...654A..41R,2024A&A...686A..44R,2025A&A...695A.240H} and references therein. The two types of emissions are generally related as $I_{R} \propto I_{X}^{k}$ \citep{2001A&A...369..441G}. This correlation probes the relative spatial distributions of the thermal and non-thermal components, where a slope k > 1 indicates that the non-thermal emission has a narrower distribution than the thermal emission. Conversely, a sublinear k < 1 indicates a broader non-thermal distribution, supporting a scenario where cosmic ray particles are accelerated and transported by turbulence generated on large scales.

We use the uGMRT Band-3, MeerKAT UHF-band, ASKAP 887 MHz, and MeerKAT L-band images to investigate how various quantities in these two regimes correlate spatially and spectrally. The analysis is automated with PT-REX \citep{2022NewA...9201732I}, using \textit{CASA} \citep{2022PASP..134k4501C} to measure the surface brightnesses and \textit{LinMix} \citep{2007ApJ...665.1489K} to perform the fitting. We chose \textit{LinMix} because it uses a Bayesian statistical approach to account for the errors in the measurements and the intrinsic scatter of the linear regression. First, we mask out the positions of all contaminating radio and X-ray sources in all bands. We then generate a grid of square cells over the 3$\sigma$ MeerKAT detection regions, where each cell is the size of the radio beam in area - see Figure \ref{fig:ptp-sb}. The same grid is used in all radio bands to study how the measured correlations change across frequency. Three distinct regions are identified to be able to study localised trends, i.e. inwards/outwards from the cold front, and the northern extension. For consistency, we use the same grid of cells for all studies except otherwise stated. The measurements are fit to a power-law relation in log-log space, i.e.
\begin{equation}
\log(I_{R}) = A + k \log(I_{X}),
\end{equation}
where $k$ is the slope of the correlation. Since \textit{LinMix} models the data as a mixture of Gaussian distributions, we use three Gaussians as a prior during fitting to correspond with our distinct regions identified. 

We present results for the radio surface brightness compared to the X-ray surface brightness, temperature, pseudo-pressure and pseudo-entropy. The radio and X-ray surface brightness of each cell and the associated uncertainties are calculated as in \citet{2022NewA...9201732I}. The other quantities and their uncertainties are taken as the mean in each cell corresponding to the nominal and uncertainty maps, respectively. The $I_{R}/I_{X}$ plane for each frequency is shown in Figure \ref{fig:ptp-sb}. Here, our identified regions clearly trace the intensity of the diffuse emission, with the region inside the cold front being the most intense (black points), then the region outside (red), and lastly, the northern extension (yellow). The average slope across frequency is $\langle k \rangle = 0.62 \pm 0.1$. Such a sub-linear correlation is typical for giant halos where the preferred mechanism of (re)-energisation is that of turbulent re-acceleration \citep[e.g.,][and references therein]{2014IJMPD..2330007B,2019SSRv..215...16V,2021A&A...650A..44B}. We note, however, that the mini-halo-like component at the centre of A\,2142 also shows a sub-linear slope \citep{2024A&A...686A..44R}. 

\begin{figure}
\centering
\includegraphics[width=\columnwidth]{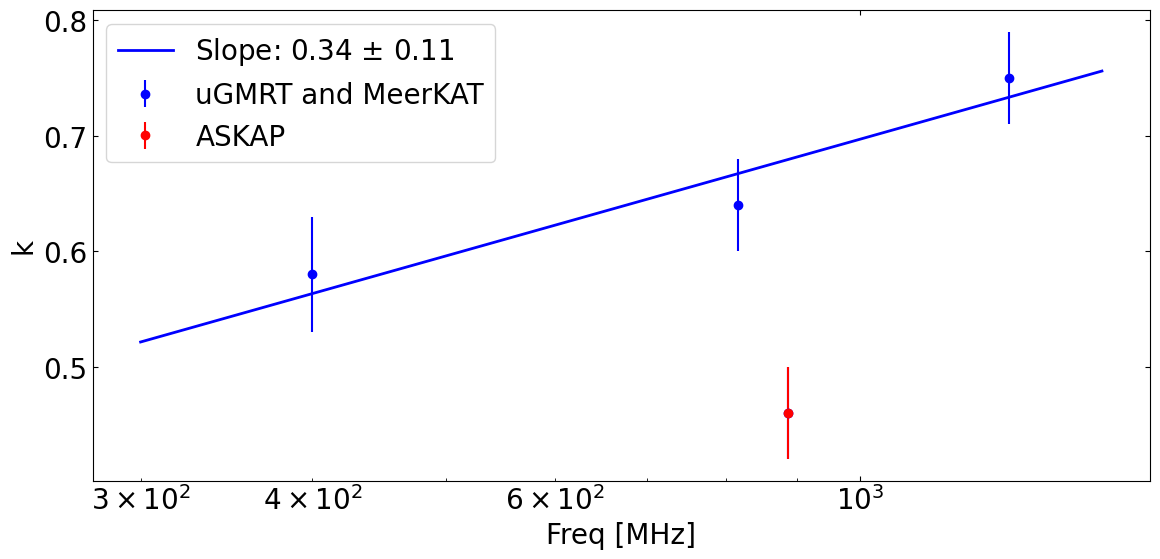}
\caption{A\,3558 $I_{R}/I_{X}$ slope versus frequency. The correlation slopes from Figure \ref{fig:ptp-sb} are plotted against frequency. The ASKAP 887 MHz data point is coloured red and excluded from linear fitting.}
\label{fig:ptp-sb-vs-freq}
\end{figure}

\begin{figure}
\centering
\includegraphics[width=0.85\columnwidth]{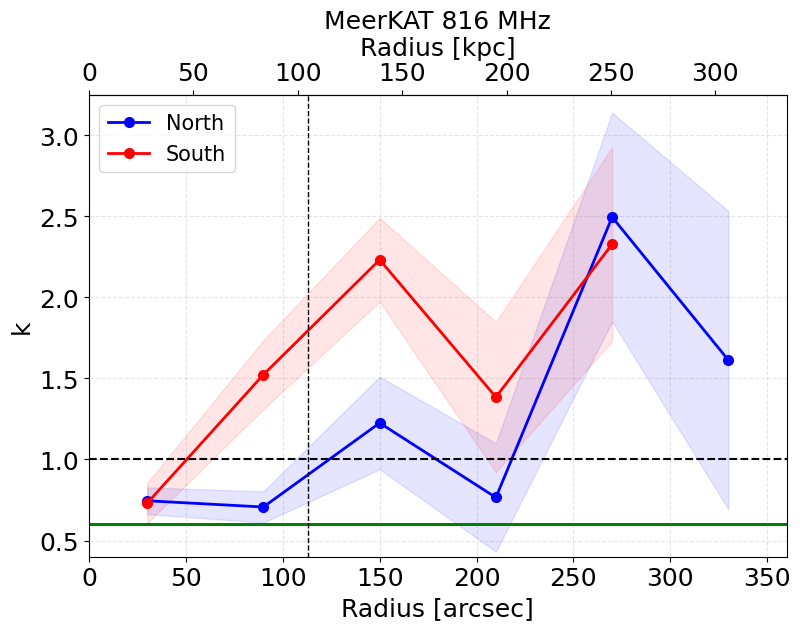}
\includegraphics[width=0.85\columnwidth]{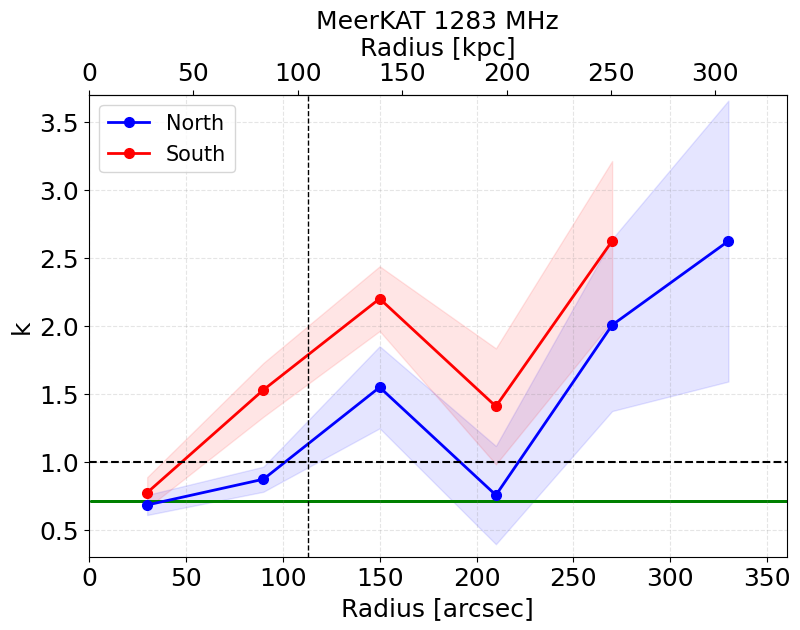}
\caption{A\,3558 $k$-correlation as a function of radius for the MeerKAT UHF-band and L-band data. The green horizontal line shows the best-fit $k$-values found in Figure \ref{fig:ptp-sb}. The black horizontal dashed line shows the unity level, and the vertical dashed line shows the radius of the northwestern cold front.}
\label{fig:ptp-sb-radial}
\end{figure}

The remaining planes are shown in Appendix \ref{sec:radio_surface_brightness_correlation_planes}. The best-fit parameters are summarised in Table \ref{tab:ptp1} along with their Pearson and Spearman coefficients. All are constrained except for uGMRT/temperature, and all show strong correlations, except for weak temperature relations. We note that the strong correlations with pseudo-pressure and pseudo-entropy are likely driven by the strong dependence of these quantities on the X-ray surface brightness as in equation \ref{eqn:1}. 

\begin{table*}
\caption{Best-fit parameter results for the radio surface brightness ($I_{R}$) point-to-point analyses against X-ray surface brightness ($I_{X}$), X-ray temperature ($X_{T}$), X-ray pseudo-pressure ($X_{p}$), and X-ray pseudo-entropy ($X_{s}$). The Pearson and Spearman coefficients ($r_{p}, r_{s}$) are also given.}
\centering
\begin{tabular}{cccccc|cccccc}
\hline
Correlation & $\nu$ (MHz) & k & A & $r_{p}$ & $r_{s}$ & Correlation & $\nu$ (MHz) & k & A & $r_{p}$ & $r_{s}$ \\
\hline 
\multirow{4}{*}{$I_{X}$} & 400 & 0.58$\pm$0.05 & -3.37$\pm$0.23 & 0.76 & 0.80 & \multirow{4}{*}{$X_{T}$} & 400 & -7.43$\pm$17.96 & -0.64$\pm$13.66 & -0.22 & -0.26 \\
& 816 & 0.64$\pm$0.04 & -3.41$\pm$0.18 & 0.84 & 0.84 & & 816 & -11.37$\pm$5.74 & 2.00$\pm$4.35 & -0.36 & -0.42 \\
& \textit{887}  & \textit{0.46$\pm$0.04} & \textit{-4.15$\pm$0.19} & \textit{0.65} & \textit{0.69} & & \textit{887} & \textit{-7.97$\pm$3.75} &\textit{-0.43$\pm$2.84} & \textit{-0.33} & \textit{-0.37} \\
& 1283 & 0.75$\pm$0.04 & -3.11$\pm$0.19 & 0.88 & 0.89 & & 1283 & -13.72$\pm$5.78 & 3.52$\pm$4.39 & -0.39 & -0.45\\

\hline 

Correlation & $\nu$ (MHz) & k & A & $r_{p}$ & $r_{s}$ & Correlation & $\nu$ (MHz) & k & A & $r_{p}$ & $r_{s}$ \\
\hline 
\multirow{4}{*}{$X_{p}$} & 400 & 1.38$\pm$0.10 & -3.92$\pm$0.17 & 0.79 & 0.83 & \multirow{4}{*}{$X_{s}$} & 400 & -1.41$\pm$0.14 & -2.88$\pm$0.34 & -0.71 & -0.73 \\
& 816 & 1.41$\pm$0.09 & -4.21$\pm$0.15 & 0.83 & 0.83 & & 816 & -1.62$\pm$0.11 & -2.73$\pm$0.27 & -0.81 & -0.79 \\
& \textit{887} & \textit{1.04$\pm$0.09} & \textit{-4.70$\pm$0.16} & \textit{0.70} & \textit{0.72} & & \textit{887} & \textit{-1.25$\pm$0.11} & \textit{-3.47$\pm$0.26} & \textit{-0.70} & \textit{-0.71} \\
& 1283 & 1.68$\pm$0.09 & -4.00$\pm$0.15 & 0.87 & 0.88 & & 1283 & -1.93$\pm$0.12 & -2.23$\pm$0.29 & -0.85 & -0.85 \\
\hline
\end{tabular}
\label{tab:ptp1}
\end{table*}

Interestingly, if we ignore the ASKAP measurements, which is our lowest-fidelity image and was not reprocessed, the correlation slopes exhibit a completely monotonic relationship with frequency as seen in Figure \ref{fig:ptp-sb-vs-freq}, with a slope of 0.34 $\pm$ 0.11 in log-linear space. This trend is most likely due to the (intrinsically negative) slope of the synchrotron spectrum. The northern extension seems to manifest trends that are different from the remaining area in the surface brightness, pseudo-pressure, and possibly pseudo-entropy correlations, most obvious in the MeerKAT planes (the right-hand-side panels of Figures \ref{fig:ptp-sb}, \ref{fig:I_R_X_p}, \ref{fig:I_R_X_s}). To investigate the local trend in each of the three regions identified, independent fits of the separate regions show that inside and outside the cold front have consistent slopes, which are consistent with the overall slope. However, the northern extension is generally much flatter than the average, possibly even negative, depending on where the boundary is drawn. If the northern extension is ignored, the remaining regions exhibit a combined fit that is slightly steeper but still remains sub-linear and monotonic with frequency. The corresponding slopes in this case are k = 0.78 $\pm$ 0.12, 0.86 $\pm$ 0.05, 0.69 $\pm$ 0.06, 0.94 $\pm$ 0.05 at 400 MHz, 816 MHz, 887 MHz and 1283 MHz, respectively.

Recent works \citep[e.g., ][]{2023A&A...678A.133B,2024A&A...686A...5B} have presented the radial trend of $k$ by taking:
\begin{equation}
k(r) = \frac{\Delta(\log I_R)}{\Delta(\log I_X)},
\label{eq:placeholder}
\end{equation}
where $\Delta(\log I)$ is the difference in logarithm of the surface brightnesses of two consecutive annuli. Such analysis reveals which type of emission is changing more/less rapidly at a particular distance throughout the cluster volume. We use the same sectors as before, but now spaced by 60$^{\prime\prime}$ to increase signal-to-noise. For clarity, in Figure \ref{fig:ptp-sb-radial}, we show these profiles only for the MeerKAT UHF- and L-band data, as these have the highest signal-to-noise. The profiles start sublinear, being consistent with the aggregate value from the point-to-point analyses, suggesting that the X-ray emission is more peaked (the gradient is higher) than the radio around the cluster centre. Thereafter, the south becomes superlinear and remains as such, suggesting that beyond the centre, the radio emission changes more rapidly than the X-ray. Note the slight dip around $r\sim200^{\prime\prime}$ (185 kpc), suggesting a slightly shallower gradient in the radio at this location. On the other hand, the north remains sublinear up until the X-ray cold front and then proceeds with a similar behaviour to the south, i.e., an increase to linear/superlinear with a drop around $r\sim200^{\prime\prime}$. The weighted mean $k$-values in the northern direction for UHF- and L-band are 0.77 $\pm$ 0.06 and 0.80 $\pm$ 0.06, respectively, and in the south are 1.16 $\pm$ 0.10 and 1.18 $\pm$ 0.09, respectively. These opposing trends illustrate the complex brightness profiles of both the radio and X-ray diffuse emissions, highlighting that a constant $k$-value does not accurately represent the entire extent.

\begin{figure}
\centering
\includegraphics[width=\columnwidth]{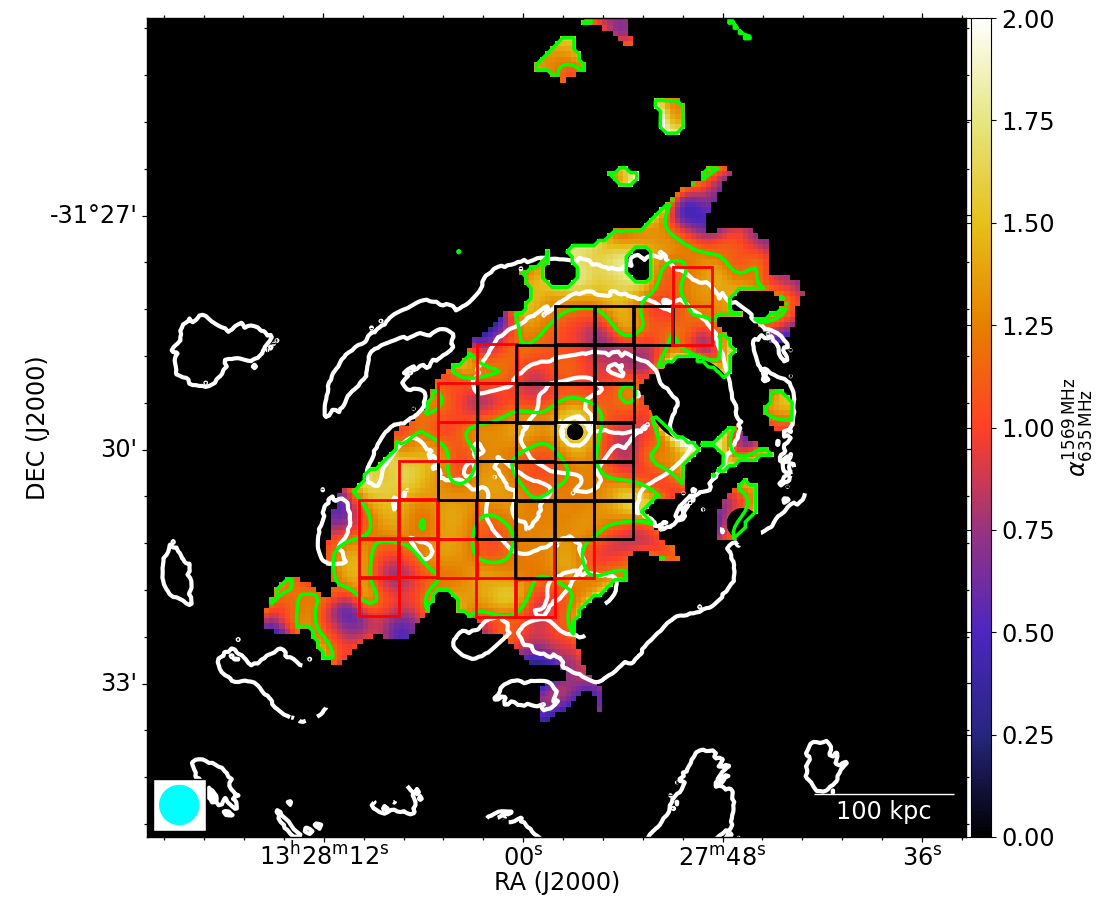}
\includegraphics[width=0.9\columnwidth]{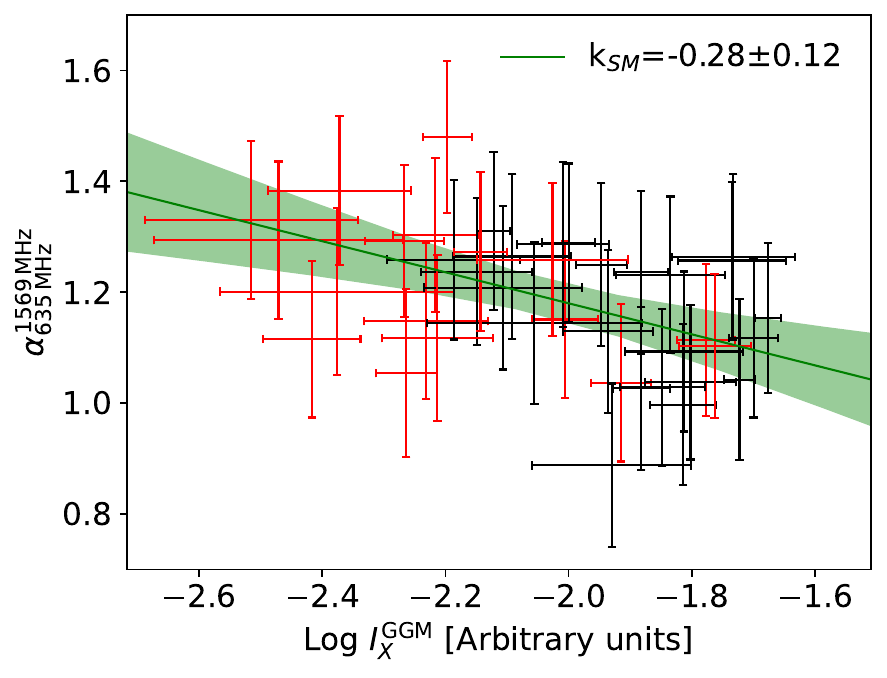}
\caption{A3558 $\alpha/I_{X}^{GGM}$ correlation. \textit{Top:} 30$^{\prime\prime}$ resolution spectral index map with a subset of the mesh grid from Figure \ref{fig:ptp-sb} overlaid. A green contour shows the 1.2 spectral index level and white contours show the X-ray GGM intensity \textit{Bottom:} color-coded $\alpha/I_{X}^{GGM}$ correlation.}
\label{fig:ptp-alpha}
\end{figure}

\subsection{Point-to-point analysis: spectral index}

We also perform the same analyses as above for the radio spectral index map, but now in log-linear space as
\begin{equation}
\alpha = A + k \log(I_{X}).
\end{equation}
Similar to the X-ray quantities, the measurement of the spectral index and its associated uncertainty are taken as the mean of each cell in the nominal and uncertainty maps, respectively. We find that the spectral index distribution at 30$^{\prime\prime}$ resolution does not correlate with any of the X-ray surface brightness or thermodynamic maps. 
The only quantity here that gives a constrained fit is against the X-ray GGM map - where the GGM measurements were taken as the mean and standard deviation of each cell. This correlation is expected given the relations shown in Figure \ref{fig:radial_profiles_alpha}. Due to the reduced area of significance, we have about half the data points of the surface brightness analysis, but still enough to make the comparisons statistically meaningful. The fit is shown in Figure \ref{fig:ptp-alpha}, with $k=-0.28\pm0.12$, $A=0.62\pm0.25$, and correlation coefficients of $r_{p}=-0.45$ and $r_{s}=-0.50$. We also used PT-REX's Markov Chain Monte Carlo algorithm to verify this result, shifting the sampling grid by a random number of pixels each iteration, finding a consistent fit. This mild anti-correlation, similar to the radial profile in Section \ref{sec:radial_profiles1}, suggests a direct link between the thermal ICM processes responsible for the X-ray cold front traced by the GGM map and the local (re)-energisation of the non-thermal component. 




\section{Discussion}
\label{sec:discussion}

Our detailed analysis of the diffuse emission in A\,3558 clearly shows the difficulty in its classification and highlights the complexity of the interplay between the thermal and non-thermal components in galaxy clusters. The results can be summarised as follows:
\begin{enumerate}
\item the diffuse emission is very faint with a radio power of $P_{\rm 1.4\,GHz} = 6.8\pm0.9 \times 10^{22}$ W/Hz, an LLS of $\sim$550 kpc and spectral index $\alpha_{\rm 400\,MHz}^{\rm 1500\,MHz} = 1.18 \pm 0.10$;
\item the northern and southern regions have some properties in common and some unique properties: both roughly follow an exponential radial profile (a double in the north at UHF and a single elsewhere) with local deviations and both have positive spectral curvature at their ends; however, the north is extended with a steeper spectrum and is dominated by radio emission, while the south is truncated with a flatter spectrum and dominated by X-ray emission;
\item the spectral index of the diffuse emission is generally anti-correlated with the X-ray GGM and residual maps, in particular, a strip of flatter spectrum in the north is coincident with the peak of the X-ray GGM map and just inside the cold front; 
\item the slope of the I$_{R}$-I$_{X}$ correlation is sublinear, steepening linearly with observing frequency, and so do the correlations with respect to the other thermodynamic quantities. The radial trend starts as sublinear and then increases to linear/superlinear in the outer regions.
\end{enumerate}

Here, we discuss the results in the broader context of the formation of the diffuse emission and the merger history of the cluster. We try to (1) classify the diffuse emission, and (2) understand its origin in the context of the ongoing merger and accretion in the SSc core.

\subsection{Classifying the diffuse emission}

The strict classification of diffuse radio emission in galaxy clusters into giant halos, which occur in merging clusters, and mini-halos, which occur in cool-core clusters, is being challenged by the discovery of an increasing number of anomalous sources. For example, the obvious case of CL1821+643, which is a cool-core cluster but hosts a giant halo \citep{2014MNRAS.444L..44B}. However, many more have been found in recent years, particularly at low frequencies, some containing \textit{hybrid} halos with an inner mini-halo-like component and outer (sometimes USS) giant-halo-like components, e.g. PSZ1G139.61+24.20, RXJ1720.1+2638 and MS 1455.0+2232 \citep{2018MNRAS.478.2234S,2019A&A...622A..24S,2021MNRAS.508.3995B,2024ApJ...961..133G}. In particular, \citet{2024A&A...686A..82B} studied these systems, as well as Abell 1068 - all cases where Mpc-scale diffuse emission is found beyond the confining cold fronts in cool-core clusters. They suggest that minor/off-axis mergers are sufficient to induce gas sloshing and related cold fronts and inject enough turbulence into the non-thermal component to produce halo-like structures, but are insufficient to fully disrupt the cool-core. Furthermore, \citet{2024ApJ...961..133G} also studied A\,3444, finding multiple cold fronts in a spiral-like configuration with faint extensions of the radio emission beyond the inner fronts but confined by the outer fronts, indicating that mini-halos can be driven by large-scale sloshing turbulence. Then there are also borderline cases, where clusters that are not clearly relaxed nor completely disturbed host transitional/intermediate halos, e.g., RXC\,J0232.2-4420 \citep{2019MNRAS.486L..80K} and SPT-CL J2031-4037 \citep{2020MNRAS.493L..28R}. One such intermediate/hybrid halo is that of Abell 2142 \citep{2017A&A...603A.125V,2023A&A...678A.133B,2024A&A...686A..44R}. Here, the diffuse emission exhibits (1) a mini-halo-like structure confined by cold fronts in the core, likely powered by turbulence from gas sloshing, (2) a ridge-like structure extending along the cluster axis aligned with a low-entropy gas trail, likely powered by larger-scale sloshing turbulence, and (3) a more extended USS halo-like structure possibly tracing older turbulent (re)-acceleration or residual merger activity and is confined by a large-scale cold front. 
As more and more examples of seemingly \textit{anomalous} cases are being found, it is clear that our binary classification convention is quickly becoming inadequate, and a broader system is required. 

\begin{figure}
\centering
\includegraphics[width=\columnwidth]{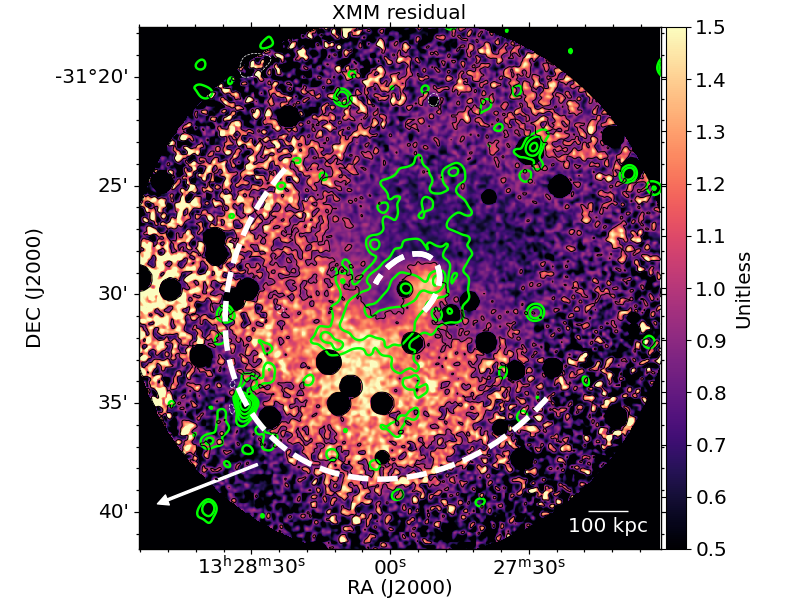}
\caption{A\,3558 X-ray residual map from Figure \ref{fig:continuum_images_xray} with a black contour at the unity level to delineate between regions of local under/over-density. The cold fronts are overlaid in white at 113$^{\prime\prime}$ \citep[105 kpc,][]{2007A&A...463..839R} NW and 9.7$^{\prime}$ \citep[540 kpc,][]{2023MNRAS.526L.124M} SE and the diffuse radio emission in green. The white arrow points in the direction of SC\,1327--0312.}
\label{fig:xray_residual}
\end{figure}

In light of the rigid paradigm and the conflicting intrinsic characteristics of the diffuse emission and those of the host cluster, the classification of the diffuse emission in A\,3558 is non-trivial. \citet{2007A&A...463..839R} highlighted that the cluster has properties both of relaxed cool-core clusters and disturbed merging clusters (see Section \ref{sec:a3558}). Interestingly, \citet{2023MNRAS.526L.124M} showed that, in addition to the one in the core, A\,3558 has a pair of large-scale cold fronts following the orientation of the cluster axis, indicating significant gas sloshing throughout the cluster volume. Figure \ref{fig:xray_residual} shows the location of the southeast front 9.7$^{\prime}$ ($\sim$540 kpc) from the surface brightness peak on the X-ray residual map and in relation to the diffuse radio emission. It is clear that the southeastern overdensity traces the large-scale cold front and compresses the diffuse radio emission. Hence, a scenario similar to A\,2142 seems to be at play, i.e., sloshing motions on small scales (inside the cooling core) and large scales (outside the core) within a dynamically intermediate cluster which is not fully relaxed nor disturbed. 

At this stage, we recall the definition of a radio mini-halo by \citet{2017ApJ...841...71G}, who state that they should have a maximum radius of 0.2R$_{500}$, i.e., the boundary where non-gravitational processes such as cooling, AGN and stellar feedback become increasingly important. The R$_{500}$ radius for A\,3558 is $\sim$1.6 Mpc \citep{2020MNRAS.497...52H}. Hence, the diffuse emission, with LLS of 550 kpc, is well within the 0.2R$_{500}$ boundary. Therefore, given the sloshing motions, the observed source size and the agreement with certain radio power scaling relations described below, we interpret the diffuse radio emission as a peculiar \textit{mini-halo} - one which leaks beyond the inner cold front to the northwest filling the negative sloshing spiral, but confined to the southeast by the outer cold front and its positive excess. As stated before, we are aware of the difficulties and limitations of such a sharp classification. However, taking into consideration the thermal and non-thermal properties, and how they relate to each other, the classification as a mini-halo is the one which best fits the observed source properties.

We note that our observed point-to-point surface brightness correlation, being $<1$, is atypical of a traditional mini-halo. However, we also show that the correlation slope increases to linear/superlinear (that typical of mini-halos) in the outer regions. We suggest that the overall sublinear slopes found in Figure \ref{fig:ptp-sb} are driven by the sublinear correlation in the core region, which is in turn driven by the radio emission being more uniform in the centre than the X-rays. In fact, the radial trend of $k$ is similar to what is seen in simulations of mini-halos, particularly Figure 6 of \citet{2015ApJ...801..146Z}, where the radio profile is shallower than the X-ray ($k<1$) at small radii and then becomes steeper at large radii ($k>1$). Nevertheless, this is not the first mini-halo to be observed with an overall sublinear correlation: \citet{2024A&A...686A..44R} show using MeerKAT L-band data that the mini-halo-like component in A\,2142 (labelled H1 in their Figure 13) has a similar slope to our L-band result, and \citet{2025A&A...695A.240H} show using uGMRT 700 MHz and MeerKAT L-band data that the mini-halo component in RX\,J1347.5--1145 also has a sublinear trend which increases with frequency.

\begin{figure*}
\centering
\includegraphics[width=0.49\textwidth]{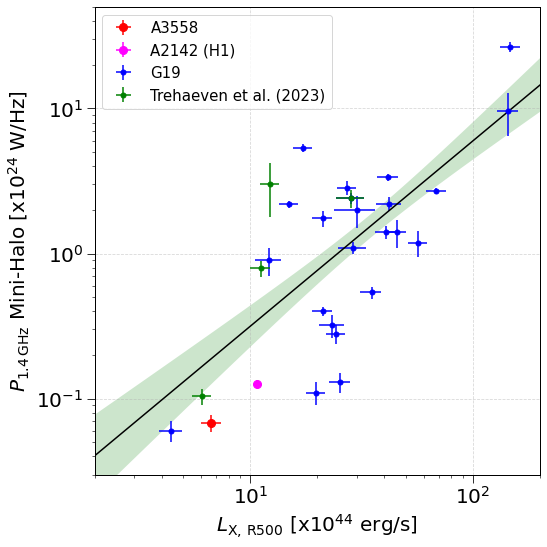}
\includegraphics[width=0.49\textwidth]{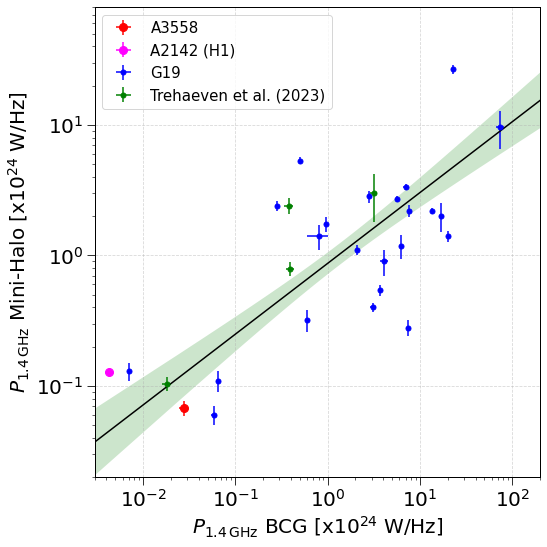}
\caption{Mini-halo radio power correlation planes against X-ray luminosity (\textit{left}) and BCG radio power (\textit{right}). Blue points are from the sample presented in \citet{2019ApJ...880...70G}, green points from \citet{2023MNRAS.520.4410T}. The A\,2142 mini-halo is marked in magenta and A\,3558 in red. The best-fit power-law is plotted with a solid black line with the 1$\sigma$ confidence band shaded in green.}
\label{fig:mini-halo-corr-planes}
\end{figure*} 

\subsection{Mini-halo scaling relations}

In light of our interpretation of the diffuse emission, it is important to examine where A\,3558 fits into known radio power correlations for mini-halos. In Figure \ref{fig:mini-halo-corr-planes}, we use our derived radio power to show the position of the A\,3558 mini-halo in the radio power versus X-ray luminosity and BCG radio power planes that mini-halos are known to follow, i.e., P$_{\rm 1.4\,GHz}^{\rm MH}$ - L$_{\rm X,\,R500}$ and P$_{\rm 1.4\,GHz}^{\rm MH}$ - P$_{\rm 1.4\,GHz}^{\rm BCG}$ \citep{2019ApJ...880...70G}. The former correlation highlights the connection between the thermal and non-thermal components of the ICM (which is strongest in the core), while the latter shows that the radio emission from the BCG \citep[always radio loud when a mini-halo is present,][]{2015A&A...581A..23K} is a likely primary supplier of fossil relativistic electrons required for (re)-acceleration mechanisms. We add the A\,2142 mini-halo into these plots for comparison. A power-law is fitted with the BCES package\footnote{\url{https://github.com/rsnemmen/BCES}} to quantify the relationships, which are found to be
\begin{equation}
\log(P_{\rm 1.4\,GHz}^{\rm MH}) = (-0.06 \pm 0.09) + (0.55 \pm 0.08)\log(P_{\rm 1.4\,GHz}^{\rm BCG}),
\end{equation}
and
\begin{equation}
\log(P_{\rm 1.4\,GHz}^{\rm MH}) = (-1.78 \pm 0.35) + (1.28 \pm 0.22)\log(L_{\rm X,\,R500}).
\end{equation}
It is clear that A\,3558 is consistent with the correlations and is one of the faintest mini-halos observed to date. The location of A\,3558 in these two plots strengthens the fact that both correlations hold down even at the faint end, spanning two orders of magnitude in $L_{X}$, four orders of magnitude in BCG power and more than two orders of magnitude in mini-halo power. Additionally, \citet{2022A&A...660A..81V} noted that the radio power is well below the mass/power relation (M$_{500}$-P$_{\rm 1.4\,GHz}$) of giant halos \citep{2021A&A...647A..51C}, reiterating our classification of a mini-halo. 

\subsection{Gas sloshing shaping and powering the mini-halo}
\label{sec:gas_sloshing}
The morphology and orientation of the mini-halo share a striking resemblance to that of the sloshing thermal gas. From the X-ray surface brightness residual map in Figures \ref{fig:continuum_images_xray} and \ref{fig:xray_residual}, we see the mini-halo extends north into a region with relatively lower thermal density and is truncated to the south by a region of higher density. The northern region is further characterised by a lower thermal pressure (Figure \ref{fig:I_R_X_p}) but higher temperature (Figure \ref{fig:I_R_X_T}) and entropy (Figure \ref{fig:I_R_X_s}), a steeper non-thermal spectrum (Figure \ref{fig:spec_index_map}) and positive spectral curvature (Figure \ref{fig:curv_index_map}). On the other hand, the south is characterised by higher pressure but lower temperature and entropy, a flatter spectrum but positive curvature as well. These trends are interpreted as the sloshing gas shaping the mini-halo by forming a local cavity into which the radio emission fills. Therefore, we suggest that the mini-halo is being shaped by this sloshing, which acts to (re)-distribute and/or (re)-accelerate seed fossil electrons, likely deposited into the cluster core by AGN feedback processes, similar to that suggested for typical mini-halos as in \citet{2019ApJ...880...70G}. Such a possibility was tested by \citet{2013ApJ...762...78Z} using magneto-hydrodynamic (MHD) simulations of a cool-core galaxy cluster interacting with a subcluster, where the sloshing motions generated turbulence and amplified magnetic fields, leading to the (re)-acceleration and (re)-distribution of seed relativistic electrons. 



The cause of the gas sloshing is likely related to the merger history of the cluster. We suggest \citep[similar to][]{2007A&A...463..839R} that the low-entropy core and tail extending southeast up to the large-scale cold front is the remnant of the A\,3558 relaxed cool-core that has been perturbed by a minor/off-axis merger, possibly with SC\,1327--312 \citep[mass ratio $\sim$5:1,][]{2018MNRAS.481.1055H}. The presence of multiple concentric cold fronts strongly supports a sloshing interpretation. Simulations show that sloshing occurs in the direction of the perturbation, and the first cold front is located on the side opposite to the interaction \citep{2024MNRAS.529..563R}. Hence, given its southeast position relative to A\,3558 and the outermost northwestern cold front detected by \citet{2023MNRAS.526L.124M} (not shown here), a minor/off-axis merger with SC\,1327--312 is supported. We suggest that the interaction did not destroy the cool-core outright but was strong enough to cause significant gas sloshing and drag the low entropy gas outward, similar to that described by \citet{2024A&A...686A..82B} for diffuse emission found beyond cool-core cold fronts. 

Following Equation 26 from \citet{2004IJMPD..13.1549G}, we can calculate a rough approximation of the average magnetic field strength within the mini-halo by assuming \textit{equipartition}. Here, we assume a filling factor and ratio between relativistic proton and electron energies of unity and the volume of the mini-halo to be that of a prolate ellipsoid with semi-major and minor axes as half the maximum and minimum lengths of the 3$\sigma$ UHF-band contour. Hence, we estimate a magnetic field strength, averaged over the mini-halo volume, of 0.34$\pm$0.01 $\mu$G. Then, from Equation 3 in \citet{2019SSRv..215...16V}, we estimate the characteristic age of electrons in the mini-halo to be $37.8\pm0.7$ Myr. If we consider the typical timescale of a merger to be $\sim$1 Gyr \citep{2019SSRv..215...16V}, then the estimated age, being two orders of magnitude lower, may suggest \textit{in-situ} particle (re)-acceleration, either in the form of hadronic collisions producing secondary electrons or via turbulent (re)-acceleration or some combination of the two. 

\subsection{The cold front as a particle accelerator}
\label{sec:cold_front}

Using radial profiles (Figure \ref{fig:radial_profiles_alpha}) and a point-to-point analysis (Figure \ref{fig:ptp-alpha}), we have shown a clear anti-correlation between the spectral index of the radio mini-halo and the X-ray GGM and residual maps. In particular, we have identified a strip of flatter spectrum emission $\sim$75$^{\prime\prime}$ (70 kpc) to the north just inside the cold front that is traced by the peak of the X-ray GGM map. To explain this, we propose that the cold front acts to inject energy into the ICM on the near-side of the front, where the X-ray gradient is greatest, thus affecting the local non-thermal properties of the cluster, either compressing and amplifying the local magnetic field and/or (re)-accelerating the population of local fossil electrons. We also identified a strip of ultra-steep spectrum emission ($\alpha \sim 2$) beyond the cold front at $\sim$ 250$^{\prime\prime}$ (232 kpc) where the X-ray GGM and residual maps reach minima. This is interpreted as aged/weaker turbulence and/or magnetic field relaxation linked to the relaxation/diffusion of the thermal gas in the local cavity. High-resolution polarimetry is required to understand the magnetic field orientation throughout the mini-halo area. 

\section{Conclusions and future prospects}
\label{sec:conclusions}

We presented a comprehensive multi-frequency radio and X-ray study of the diffuse emission in the central region of the galaxy cluster Abell 3558, leveraging new MeerKAT UHF-band and uGMRT Band-3 observations alongside archival ASKAP and MeerKAT L-band data, complemented by XMM-Newton X-ray imaging. We conclude that the diffuse radio emission in A\,3558 can be classified as a peculiar mini-halo. The highlights of our study and the evidence in support of our conclusions are detailed below.
\begin{enumerate}
    \item Discovery of a northern extension: We report the first detection of a faint northern extension of the diffuse radio emission beyond the innermost cold front, increasing the projected largest linear size of the emission to $\sim$550 kpc. This feature was not detected in earlier works and points to a more complex and extended mini-halo structure than previously understood.
    \item High-fidelity spectral characterisation: The diffuse emission has an integrated spectral index of $\alpha_{\rm 400\,MHz}^{\rm 1569\,MHz}=1.18\pm0.10$. This value, flatter than the ultra-steep spectrum previously inferred from ASKAP and MeerKAT L-band data alone, highlights the importance of broad and consistent frequency coverage in spectral analyses.
    \item Local spectral variations: The spectral index map shows a region of flatter spectrum ($\alpha \lesssim 1$) just inside the X-ray cold front, indicative of local turbulent (re)-acceleration, and beyond that a region of steeper spectrum ($\alpha \sim 2.0$), filling an X-ray cavity of higher temperature and entropy but lower pressure. The northern and southern edges exhibit positive curvature, suggesting electron ageing. These trends support turbulent (re)-acceleration scenarios and reveal spatially varying non-thermal processes shaped by ICM dynamics.
    \item Strong radio–X-ray correlations: The point-to-point surface brightness correlation between the radio and X-ray emission is sub-linear across all frequencies ($\langle k \rangle = 0.62 \pm 0.1$), unusual for mini-halos, supporting a turbulent re-acceleration origin. The correlation begins sublinear in the core and then increases to linear/superlinear in the outer regions. Moreover, the spectral index correlates negatively with X-ray residuals and GGM features, linking non-thermal energetics directly to gas sloshing signatures and thermal structure.
    \item Classification as a mini-halo: The prevalence of sloshing features, the spatial confinement of the emission within a (large-scale) cold front, the observed size and radio power correlations all point toward a radio mini-halo classification. However, the northern extension beyond the innermost cold front and the sublinear point-to-point correlation make it a \textit{peculiar} object and suggest that a broader classification scheme is necessary to describe it absolutely. This is only the third case of a mini-halo having such a sublinear correlation and highlights the complex nature of these sources and of their production mechanisms.
    \item Minor merger in the SSc core: The above findings, with an estimated age of relativistic electrons of $\sim$ 40 Myr, all suggest that the mini-halo in A\,3558 is powered in-situ by sloshing-induced turbulence, reinforcing the picture of an ongoing minor merger with the group SC\,1327--312 (mass ratio $\sim$5:1).
\end{enumerate}

The present paper adds another piece of information to our study of the merger activity in the Shapley Supercluster with state-of-the-art radio facilities \citep{2022A&A...660A..81V,2022ApJ...934...49G,2024arXiv240814142D,2024MNRAS.533.1394M}. It also contributes to the growing body of work exploring the role of merger-driven turbulence and gas sloshing in powering radio mini-halos. The observed spectral variations and correlations between radio and X-ray features highlight the complexity of energy transfer processes within the cluster environment. These results also place A\,3558 within a broader context of merging systems such as A\,2142, showcasing how minor mergers can generate and sustain non-thermal emission on significant spatial scales.

Looking ahead, deeper observations with the next generation of radio telescopes will be essential to further unravel the nature of mini-halos and their link to cluster dynamics. The Square Kilometre Array (SKA) represents a particularly promising avenue for future study. With its improved sensitivity and frequency coverage — especially in SKA-Mid Bands 1 and 2 and SKA-Low — it will enable a more precise characterisation of the particle (re)-acceleration mechanisms at play, as well as a comprehensive mapping of magnetic field structures within A\,3558. Additionally, high-resolution polarimetric studies will be crucial in discerning the alignment and amplification of magnetic fields, shedding further light on their role in shaping the observed radio properties.

By expanding our observational and theoretical frameworks, we will be better equipped to refine the classification of diffuse radio emission in galaxy clusters, bridging the gap between mini-halos, giant radio halos, and other hybrid structures. A\,3558 serves as a key laboratory for these investigations, offering new insights into the evolutionary pathways of clusters within the cosmic web.

\section*{Acknowledgements}

The authors thank the anonymous referee for the helpful comments
which improved the clarity of the paper. The MeerKAT telescope is operated by the South African Radio Astronomy Observatory, which is a facility of the National Research Foundation, an agency of the Department of Science and Innovation. The financial assistance of the South African Radio Astronomy Observatory (SARAO) towards this research is hereby acknowledged (www.sarao.ac.za). KT acknowledges financial support from the South African Department of Science and Innovation's National Research Foundation under the ISARP RADIOMAP Joint Research Scheme (DSI-NRF Grant Number 150551). KT, TV, GB and PM acknowledge partial support from the INAF mini-grant 2022 ShaSEE (Shapley Supercluster Exploitation and Exploration). The research of OS is supported by the South African Research Chairs Initiative of the DSTI/NRF (grant No. 81737). GDG acknowledges support from the ERC Consolidator Grant ULU 101086378. MR acknowledges support from INAF (Ricerca Fondamentale GO grant) and Prin-MUR 2022, supported by Next Generation EU (n.20227RNLY3). SG acknowledges that the basic research in radio astronomy at the Naval Research Laboratory is supported by 6.1 Base funding. This work was supported in part by the Italian Ministry of Foreign Affairs and International Cooperation, grant number ZA23GR03. SPS would like to acknowledge the financial support from SARAO. RK acknowledges the support of the Department of Atomic Energy, Government of India, under project no. 12-R\&D-TFR-5.02-0700 and the Science and Engineering Research Board (SERB) Women Excellence Award WEA/2021/000008. KK acknowledges financial support from NRF/SARAO grant UID 97930.

\section*{Data availability}
The raw MeerKAT data underlying this article is available on the SARAO archive, at \url{https://archive.sarao.ac.za}. The raw uGMRT data underlying this article is available on the uGMRT archive, at \url{https://naps.ncra.tifr.res.in/goa/data/search}. 

\bibliographystyle{mnras}
\bibliography{main}

\appendix

\section{BCG Flux density measurements and integrated spectrum}

\begin{figure}
\centering
\includegraphics[width=\columnwidth]{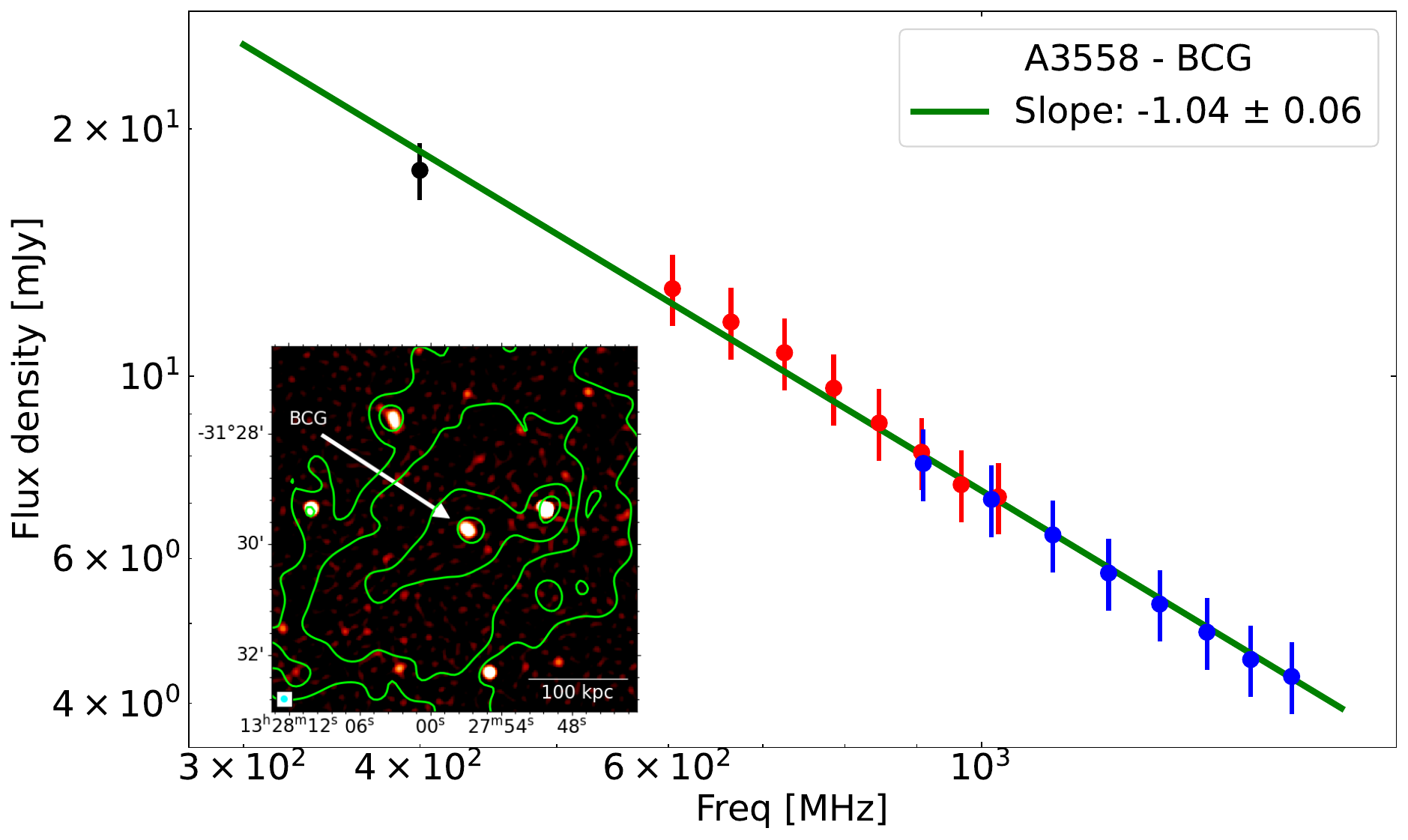}
\caption{A3558 BCG integrated spectrum. Red and blue markers show the MeerKAT UHF- and L-band data points, respectively. The black marker shows the uGMRT 400 MHz measurement. The inset is a cutout of the UHF high-resolution image from which measurements are listed in Table \ref{tab:bcg_flux_measurements}.}
\label{fig:bcg_int_spec}
\end{figure}

Below, in Table \ref{tab:bcg_flux_measurements}, we list the A3558 BCG flux density measurements from the MeerKAT HR subband images, complemented with the measurement from the uGMRT HR MFS image. The resulting spectrum is given in Figure \ref{fig:bcg_int_spec}.

\begin{table}
\caption{Flux density measurements of the A3558 BCG used in this work for the determination of its spectral index and 1.4 GHz radio power.}
\centering
\begin{tabular}{ccc}
\hline
$\nu$ & Array & Flux density \\
(MHz) &       & (mJy)        \\
\hline
400 & uGMRT (Band-3)& 17.81$\pm$1.42\\
604 & MeerKAT (UHF-Band)& 12.78$\pm$1.28\\
665 & MeerKAT (UHF-Band)& 11.64$\pm$1.16\\
725 & MeerKAT (UHF-Band)& 10.68$\pm$1.07\\
786 & MeerKAT (UHF-Band)& 9.67$\pm$0.97\\
846 & MeerKAT (UHF-Band)& 8.77$\pm$0.88\\
907 & MeerKAT (UHF-Band)& 8.08$\pm$0.81\\
967 & MeerKAT (UHF-Band)& 7.38$\pm$0.74\\
1028 & MeerKAT (UHF-Band)& 7.13$\pm$0.71\\
909 & MeerKAT (L-Band)& 7.83$\pm$0.78\\ 
1016 & MeerKAT L-Band)& 7.08$\pm$0.71\\ 
1123 & MeerKAT (L-Band)& 6.41$\pm$0.64\\ 
1230 & MeerKAT (L-Band)& 5.76$\pm$0.58\\ 
1337 & MeerKAT (L-Band)& 5.28$\pm$0.53\\ 
1444 & MeerKAT (L-Band)& 4.88$\pm$0.49\\ 
1551 & MeerKAT (L-Band)& 4.52$\pm$0.45\\ 
1658 & MeerKAT (L-Band)& 4.31$\pm$0.43\\ 
\hline
\end{tabular}
\label{tab:bcg_flux_measurements}
\end{table}

\section{Spectral index uncertainty maps}

\begin{figure}
\centering
\includegraphics[width=\columnwidth]{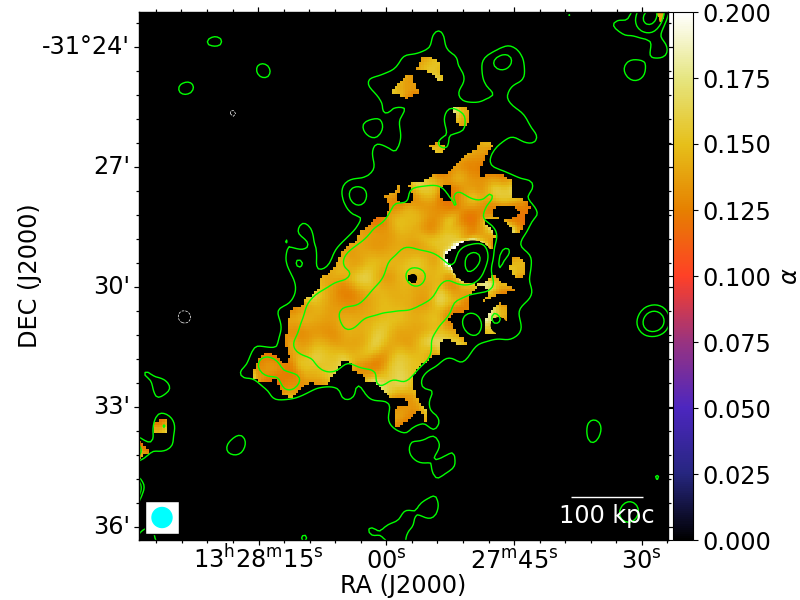}
\includegraphics[width=\columnwidth]{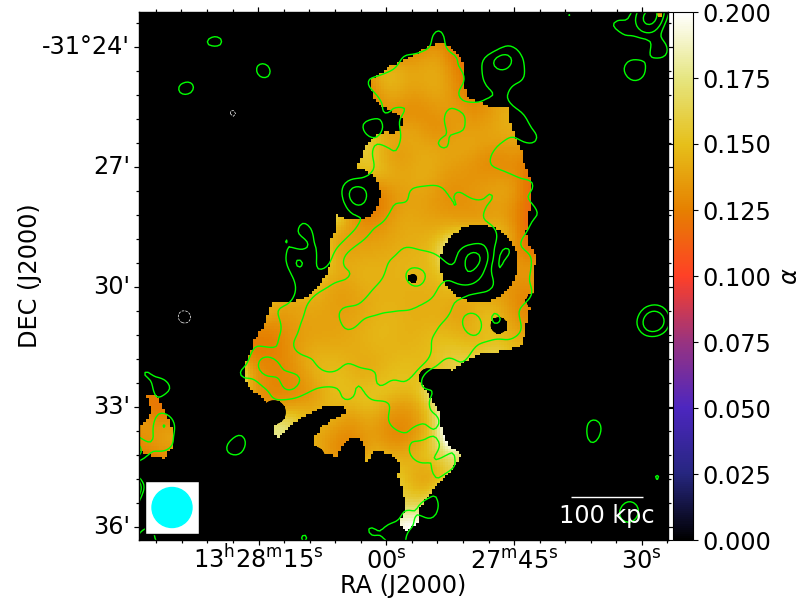}
\caption{A3558 spectral index uncertainty maps at 30$^{\prime\prime}$ resolution \textit{(top)} and 60$^{\prime\prime}$ \textit{(bottom)} with UHF-band contours overlaid.}
\label{fig:spec_index_uncert}
\end{figure}

In Figure \ref{fig:spec_index_uncert}, we show the uncertainty maps for the spectral index maps shown in Figure \ref{fig:spec_index_map} of the main body of text.

\section{Relative ICM profiles}

\begin{figure}
\centering
\includegraphics[width=0.8\columnwidth]{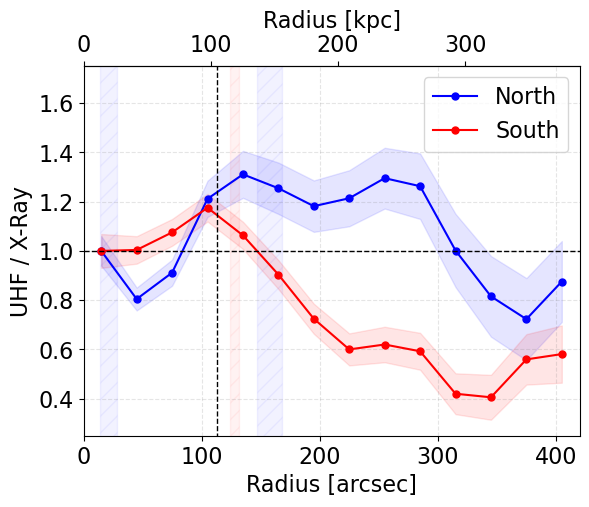}
\includegraphics[width=0.8\columnwidth]{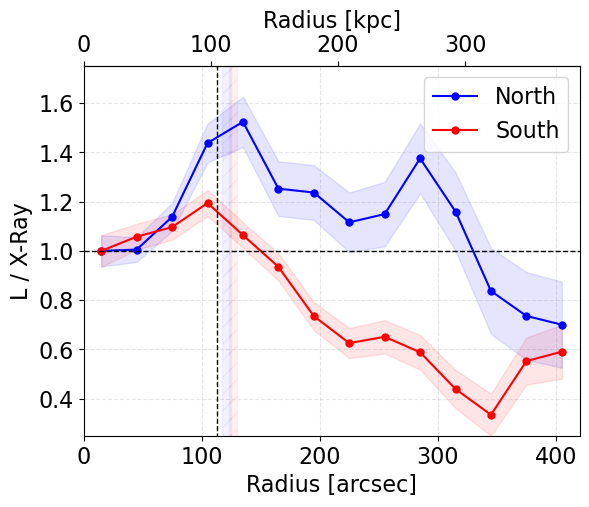}
\caption{A3558 UHF- and L- band surface brightness profiles normalised with respect to the X-ray profiles. \textit{Top:} UHF-band. \textit{Bottom:} L-band}
\label{fig:radial_profiles_sb_x}
\end{figure}

We examined whether the thermal or the non-thermal emission dominates in a given region of the diffuse radio emission in A\,3558 by analysing the trend of the ratio between the normalised radio profile and the normalised X-ray profile. Figure \ref{fig:radial_profiles_sb_x} shows the radial trend of this ratio along the same north/south directions as in Section \ref{sec:spx}. At small radii, the ratio is $>1$, except for r $\lesssim$ 75$^{\prime\prime}$ (70 kpc) in UHF-band, and reaches a local peak at or just outside the X-ray cold front. Thereafter, at intermediate radii, the radio continues to dominate in the north until the ratio reaches unity at large radii. This trend is linked to the surface brightness discontinuity formed by the cold front and the deficiency beyond that clearly seen in the residual map of Figure \ref{fig:continuum_images_xray}. On the other hand, in the south, the ratio is $<1$ beyond where the cold front would be and is linked to the surplus of X-rays seen again in Figure \ref{fig:continuum_images_xray}. 

\section{Radio surface brightness correlation planes}
\label{sec:radio_surface_brightness_correlation_planes}

\begin{figure}
\centering
\includegraphics[width=0.49\columnwidth]{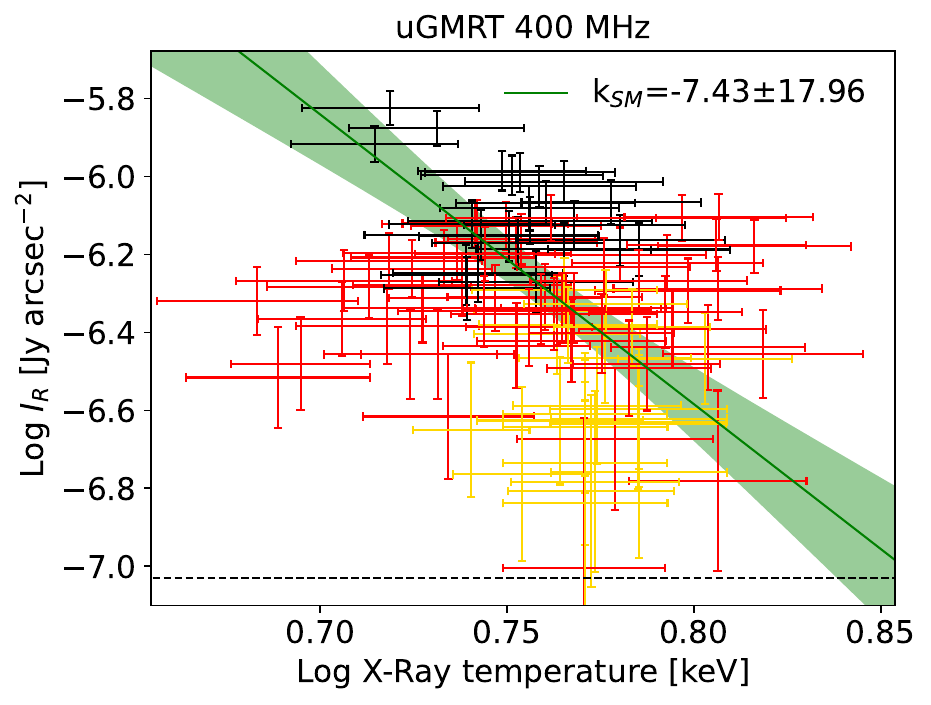}
\includegraphics[width=0.49\columnwidth]{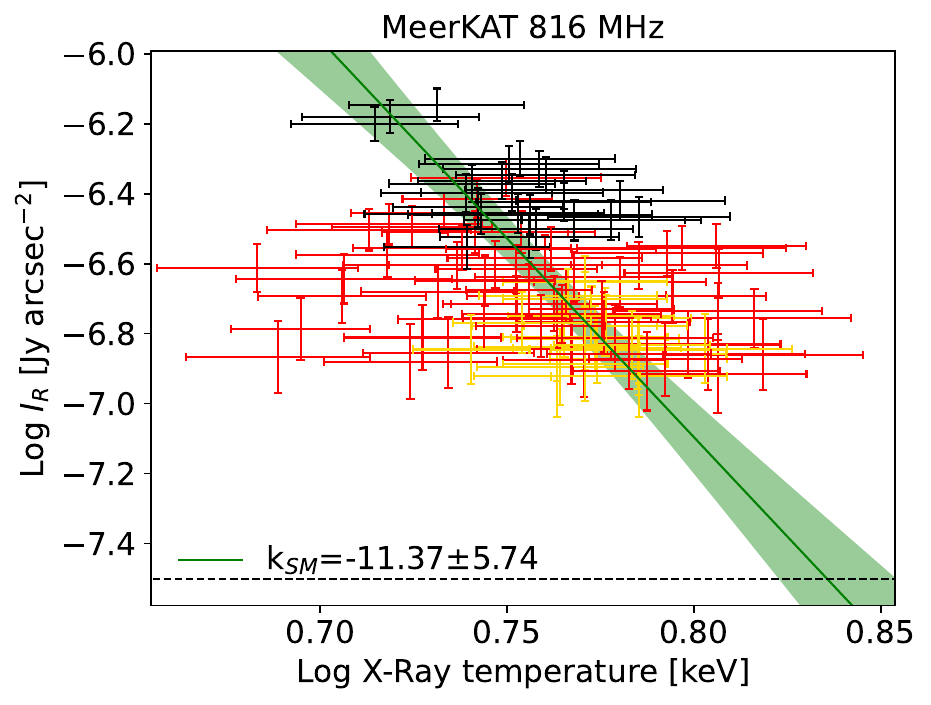}
\includegraphics[width=0.49\columnwidth]{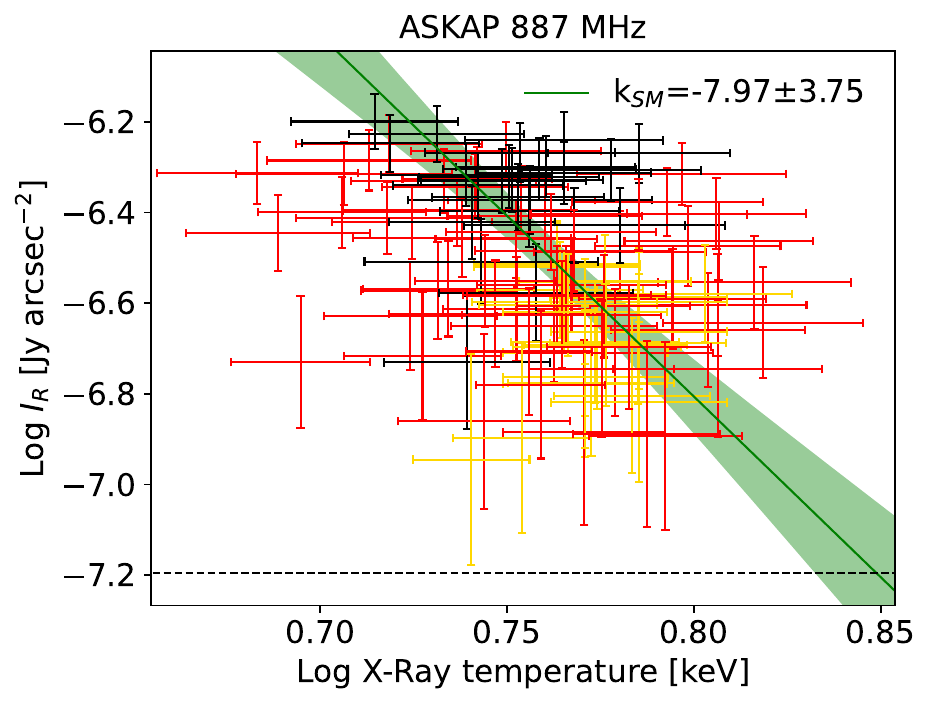}
\includegraphics[width=0.49\columnwidth]{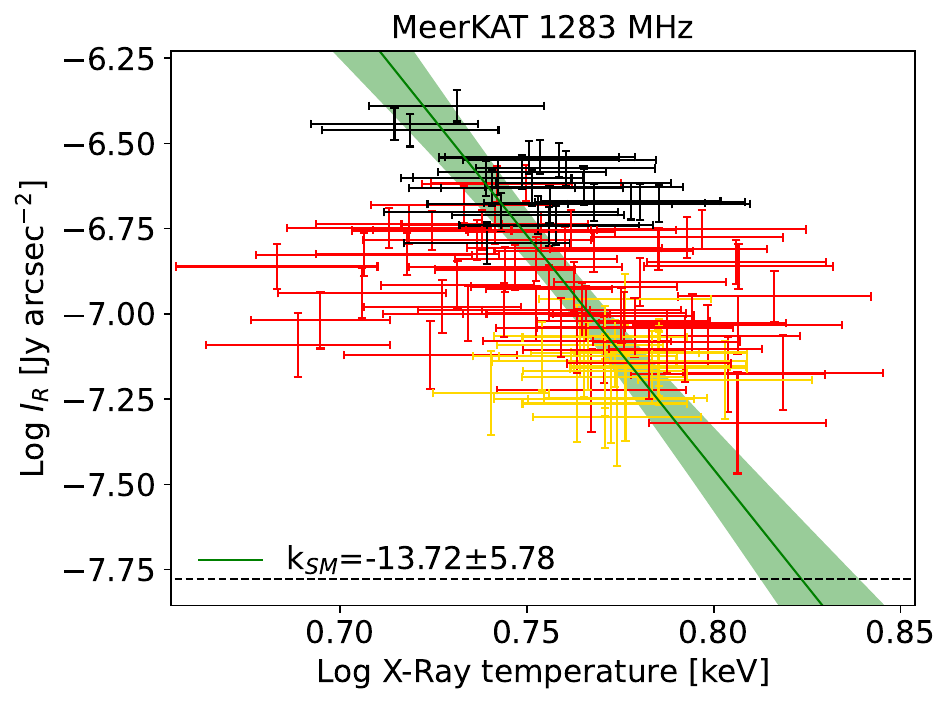}
\caption{A3558 $I_{R}/X_{T}$ planes. Similar to Figure \ref{fig:ptp-sb} but for X-ray temperature.}
\label{fig:I_R_X_T}
\end{figure}

\begin{figure}
\centering
\includegraphics[width=0.49\columnwidth]{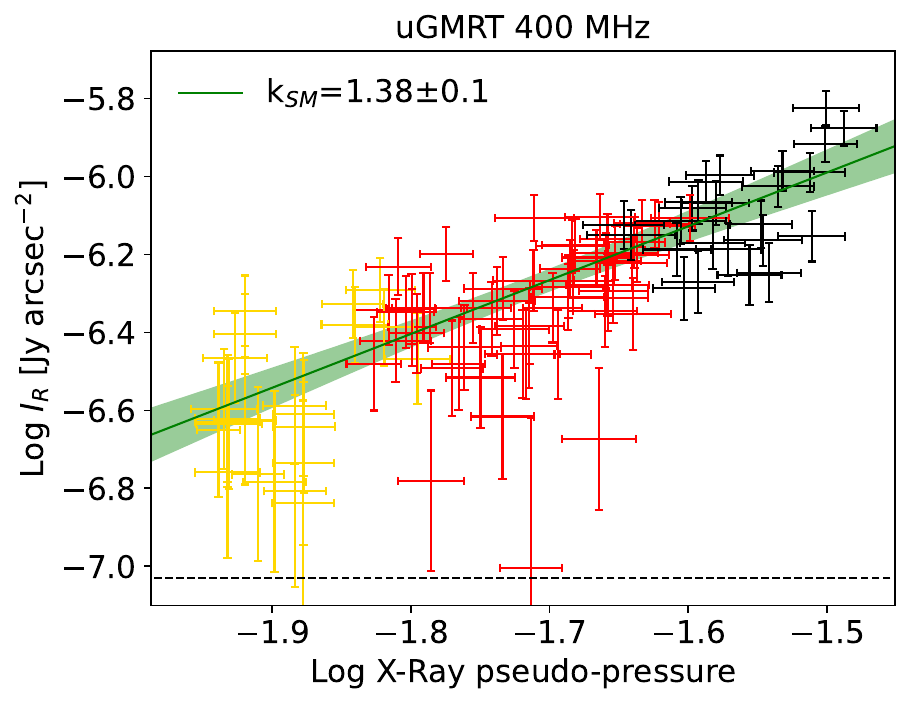}
\includegraphics[width=0.49\columnwidth]{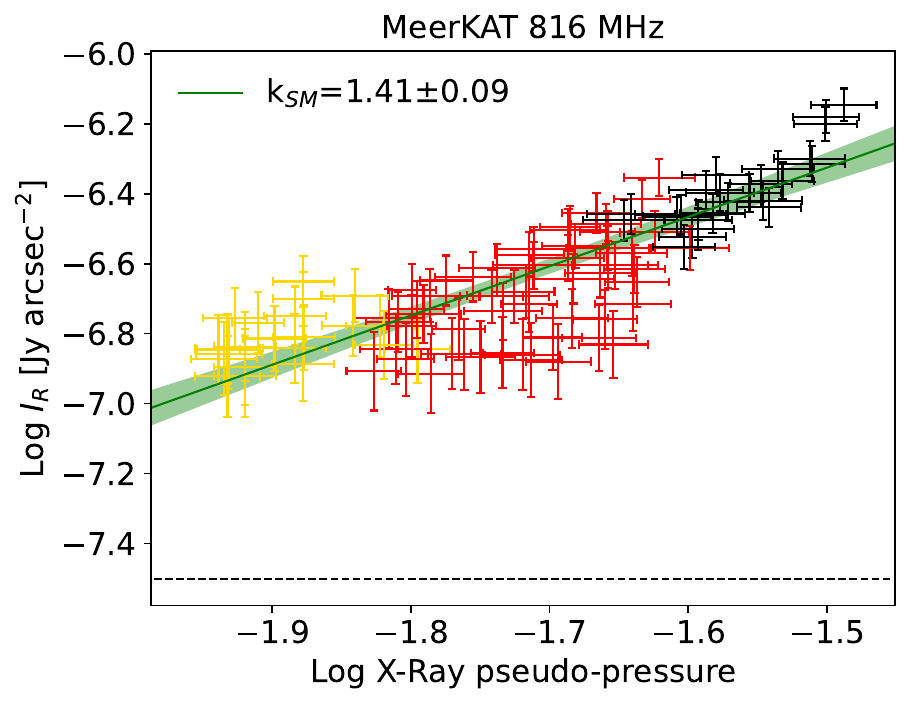}
\includegraphics[width=0.49\columnwidth]{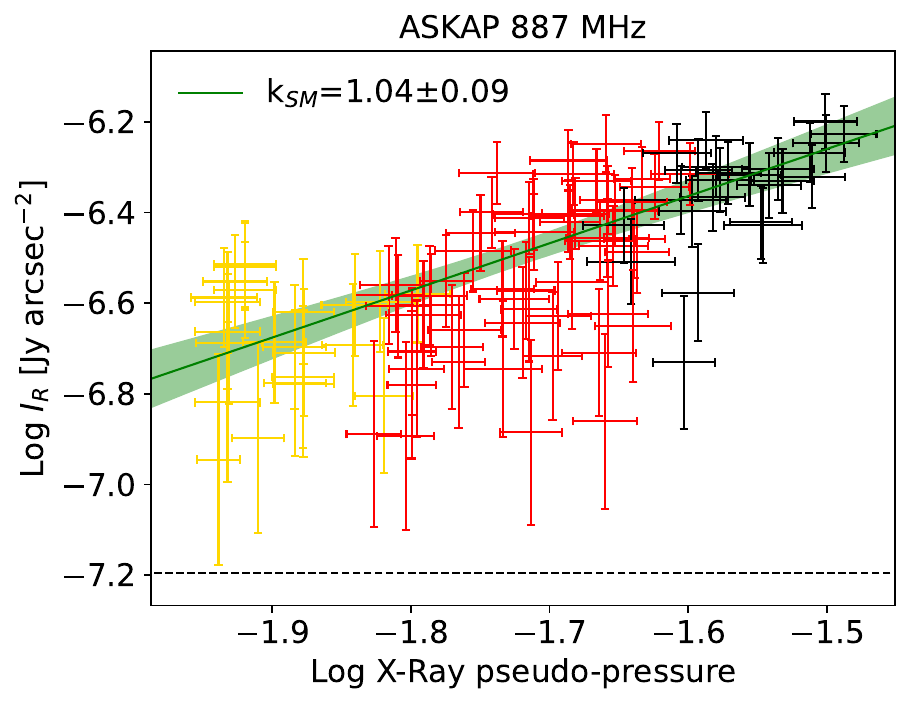}
\includegraphics[width=0.49\columnwidth]{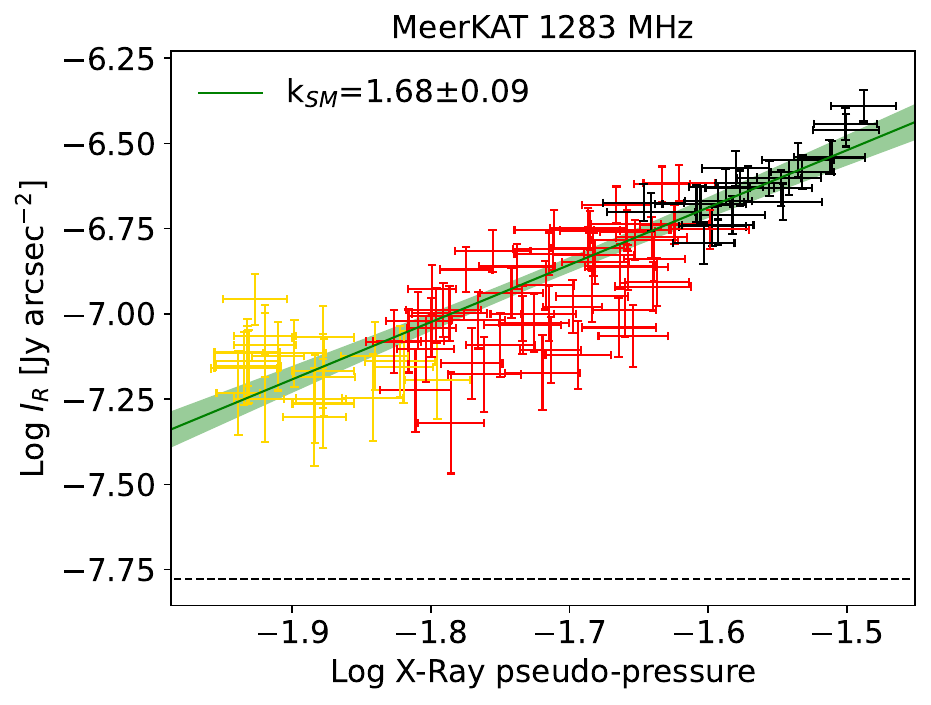}
\caption{A3558 $I_{R}/X_{p}$ planes. Similar to Figure \ref{fig:ptp-sb} but for X-ray pseudo-pressure.}
\label{fig:I_R_X_p}
\end{figure}

\begin{figure}
\centering
\includegraphics[width=0.49\columnwidth]{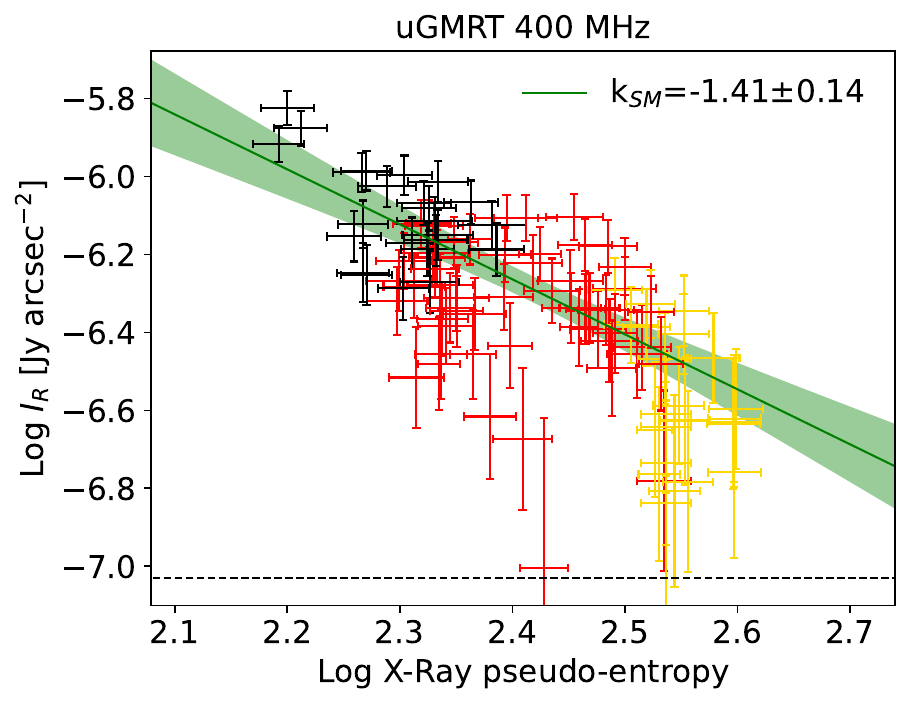}
\includegraphics[width=0.49\columnwidth]{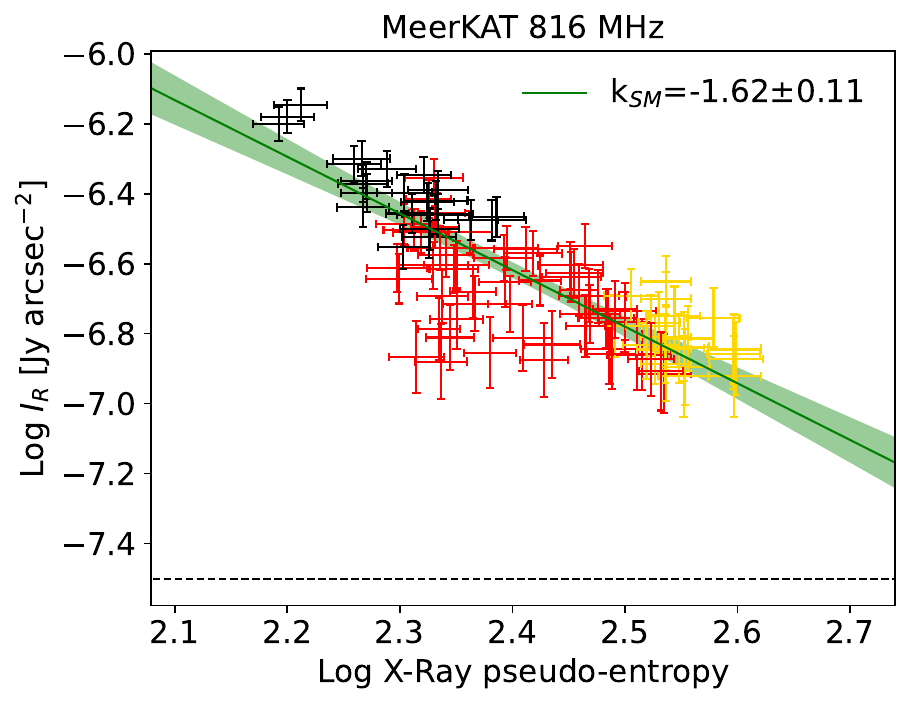}
\includegraphics[width=0.49\columnwidth]{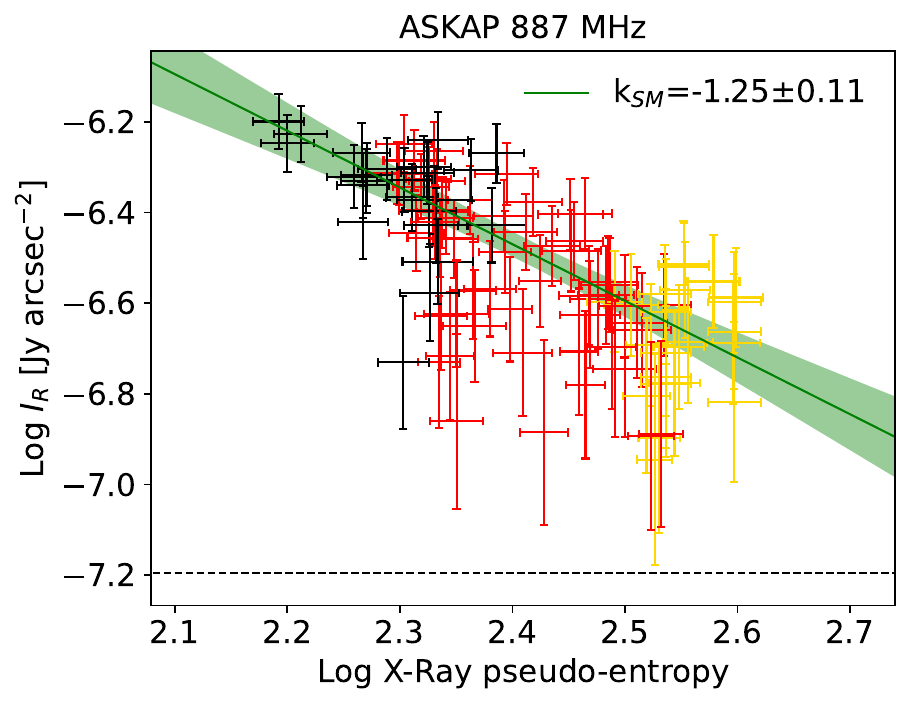}
\includegraphics[width=0.49\columnwidth]{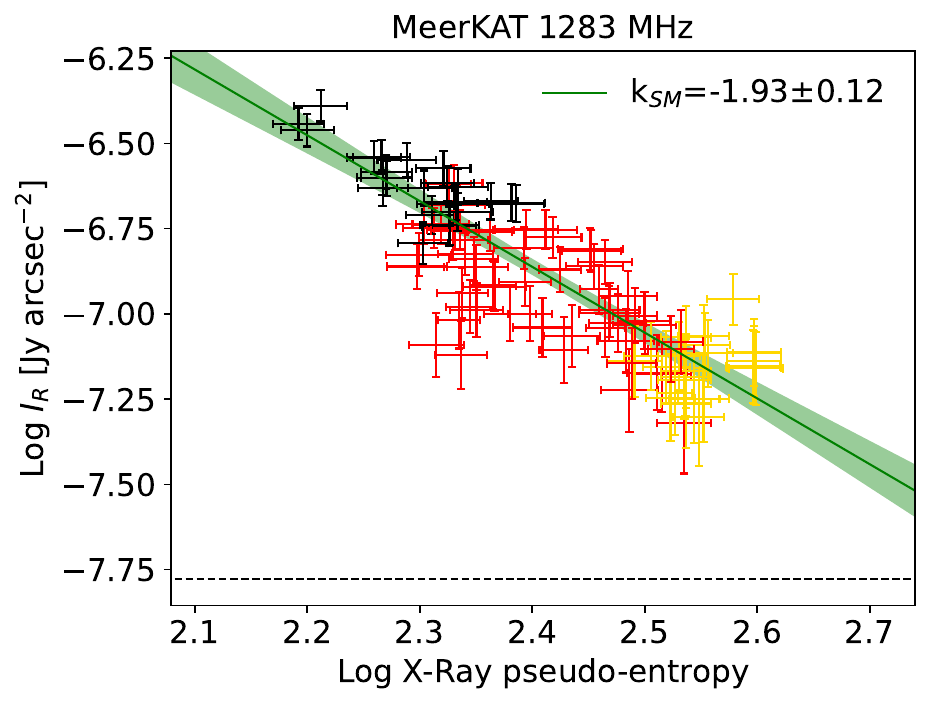}
\caption{A3558 $I_{R}/X_{s}$ planes. Similar to Figure \ref{fig:ptp-sb} but for X-ray pseudo-entropy.}
\label{fig:I_R_X_s}
\end{figure}

In Figures \ref{fig:I_R_X_T}, \ref{fig:I_R_X_p} and \ref{fig:I_R_X_s}, we show the point-to-point correlation planes between the radio surface brightness and the X-ray temperature, pseudo-pressure and pseudo-entropy, respectively.

\section{Subtracted emission}

In Table \ref{tab:subtracted_srcs}, we list all the discrete sources masked and subsequently subtracted from the diffuse emission.

\begin{table*}
\caption{Subtracted compact sources from the \textit{High resolution} UHF-band image with noise 18.5 $\mu$Jy/beam and resolution (6.6$^{\prime\prime}\times6.4^{\prime\prime}, 44.1^{\circ}$). The RA and DEC correspond to the centroid of each masked region during high-resolution imaging inside the diffuse emission area. *BCG.} 
\centering
\begin{tabular}{cc|cc}
\hline
RA, DEC (h:m:s, $^{\circ}:^{\prime}:^{\prime\prime}$) &  S$_{\rm 816\,MHz}$ (mJy) & RA, DEC (h:m:s, $^{\circ}:^{\prime}:^{\prime\prime}$) &  S$_{\rm 816\,MHz}$ (mJy)\\
\hline
13:27:50.25, -31:29:21.34 & 76.7953 $\pm$ 7.6795 & 13:27:53.15, -31:29:35.47 & 0.0283 $\pm$ 0.0028\\ 
*13:27:56.86, -31:29:43.50 & 9.8304 $\pm$ 0.983 & 13:27:54.74, -31:26:07.40 & 0.0279 $\pm$ 0.0028\\ 
13:27:54.97, -31:32:18.53 & 7.7163 $\pm$ 0.7716 & 13:28:02.50, -31:29:18.24 & 0.0272 $\pm$ 0.0027\\
13:28:02.68, -31:32:14.39 & 0.6557 $\pm$ 0.0656 & 13:27:54.94, -31:28:50.00 & 0.0267 $\pm$ 0.0027\\ 
13:27:53.10, -31:25:48.30 & 0.4631 $\pm$ 0.0463 & 13:27:48.15, -31:26:15.29 & 0.0267 $\pm$ 0.0027\\ 
13:28:12.50, -31:31:30.23 & 0.3805 $\pm$ 0.038 & 13:27:51.14, -31:31:08.06 & 0.0253 $\pm$ 0.0025\\ 
13:27:46.60, -31:27:15.72 & 0.3529 $\pm$ 0.0353 & 13:27:51.85, -31:25:37.40 & 0.025 $\pm$ 0.0025\\ 
13:28:14.24, -31:31:57.74 & 0.3309 $\pm$ 0.0331 & 13:27:51.08, -31:25:48.37 & 0.0243 $\pm$ 0.0024\\ 
13:27:56.85, -31:27:17.67 & 0.2848 $\pm$ 0.0285 & 13:27:58.08, -31:24:27.15 & 0.0241 $\pm$ 0.0024\\ 
13:27:43.68, -31:29:30.05 & 0.2787 $\pm$ 0.0279 & 13:27:56.59, -31:25:22.87 & 0.0223 $\pm$ 0.0022\\ 
13:27:52.36, -31:27:56.63 & 0.2029 $\pm$ 0.0203 & 13:28:00.01, -31:31:43.95 & 0.022 $\pm$ 0.0022\\ 
13:27:55.84, -31:28:28.06 & 0.1811 $\pm$ 0.0181 & 13:27:44.38, -31:29:45.17 & 0.0219 $\pm$ 0.0022\\ 
13:28:11.29, -31:32:17.80 & 0.1639 $\pm$ 0.0164 & 13:27:58.03, -31:32:06.05 & 0.0217 $\pm$ 0.0022\\ 
13:28:07.19, -31:31:33.80 & 0.1549 $\pm$ 0.0155 & 13:27:49.36, -31:30:57.04 & 0.0191 $\pm$ 0.0019\\ 
13:27:43.88, -31:29:59.42 & 0.1508 $\pm$ 0.0151 & 13:27:59.05, -31:29:12.12 & 0.0186 $\pm$ 0.0019\\ 
13:28:02.67, -31:28:49.05 & 0.1344 $\pm$ 0.0134 & 13:27:56.38, -31:30:46.77 & 0.0184 $\pm$ 0.0018\\ 
13:28:05.44, -31:31:33.88 & 0.1214 $\pm$ 0.0121 & 13:27:47.95, -31:26:58.45 & 0.0181 $\pm$ 0.0018\\ 
13:28:03.66, -31:30:15.63 & 0.1182 $\pm$ 0.0118 & 13:27:53.86, -31:27:24.68 & 0.0143 $\pm$ 0.0014\\ 
13:28:02.17, -31:30:55.50 & 0.1142 $\pm$ 0.0114 & 13:27:54.02, -31:32:39.13 & 0.0138 $\pm$ 0.0014\\ 
13:27:57.57, -31:33:11.94 & 0.1071 $\pm$ 0.0107 & 13:28:02.87, -31:32:00.02 & 0.0135 $\pm$ 0.0013\\ 
13:27:47.34, -31:27:30.19 & 0.1033 $\pm$ 0.0103 & 13:28:00.14, -31:31:27.79 & 0.0126 $\pm$ 0.0013\\ 
13:27:51.88, -31:32:13.48 & 0.1023 $\pm$ 0.0102 & 13:27:50.87, -31:27:20.51 & 0.0126 $\pm$ 0.0013\\ 
13:28:04.45, -31:31:58.70 & 0.1017 $\pm$ 0.0102 & 13:28:00.01, -31:29:17.59 & 0.0118 $\pm$ 0.0012\\ 
13:27:55.08, -31:30:06.89 & 0.1009 $\pm$ 0.0101 & 13:27:52.59, -31:26:36.87 & 0.0118 $\pm$ 0.0012\\ 
13:27:55.87, -31:33:32.45 & 0.0908 $\pm$ 0.0091 & 13:28:00.15, -31:32:55.79 & 0.0114 $\pm$ 0.0011\\ 
13:28:02.21, -31:31:44.55 & 0.0899 $\pm$ 0.009 & 13:28:00.94, -31:28:21.79 & 0.0113 $\pm$ 0.0011\\ 
13:28:05.21, -31:31:09.89 & 0.0787 $\pm$ 0.0079 & 13:27:50.05, -31:31:06.34 & 0.0107 $\pm$ 0.0011\\ 
13:28:01.98, -31:28:43.99 & 0.0778 $\pm$ 0.0078 & 13:27:55.07, -31:24:11.09 & 0.0104 $\pm$ 0.001\\ 
13:28:12.02, -31:32:26.44 & 0.0753 $\pm$ 0.0075 & 13:28:13.44, -31:32:28.84 & 0.0103 $\pm$ 0.001\\ 
13:27:43.44, -31:27:53.61 & 0.0721 $\pm$ 0.0072 & 13:28:10.51, -31:31:41.99 & 0.0098 $\pm$ 0.001\\ 
13:27:48.51, -31:28:45.70 & 0.059 $\pm$ 0.0059 & 13:27:50.31, -31:28:01.89 & 0.0087 $\pm$ 0.0009\\ 
13:27:44.61, -31:28:43.21 & 0.0585 $\pm$ 0.0058 & 13:27:57.30, -31:34:25.25 & 0.0086 $\pm$ 0.0009\\ 
13:27:52.51, -31:30:44.65 & 0.0522 $\pm$ 0.0052 & 13:27:54.71, -31:27:05.00 & 0.0085 $\pm$ 0.0009\\ 
13:27:56.25, -31:33:12.10 & 0.0508 $\pm$ 0.0051 & 13:27:58.74, -31:32:53.00 & 0.0066 $\pm$ 0.0007\\ 
13:28:13.88, -31:30:40.22 & 0.0478 $\pm$ 0.0048 & 13:28:15.03, -31:30:56.03 & 0.0062 $\pm$ 0.0006\\ 
13:27:50.87, -31:26:06.92 & 0.0456 $\pm$ 0.0046 & 13:27:46.92, -31:28:46.58 & 0.0061 $\pm$ 0.0006\\ 
13:27:50.71, -31:30:53.10 & 0.0422 $\pm$ 0.0042 & 13:27:56.59, -31:26:44.49 & 0.0059 $\pm$ 0.0006\\ 
13:28:04.08, -31:29:07.38 & 0.0413 $\pm$ 0.0041 & 13:27:58.32, -31:31:42.85 & 0.0057 $\pm$ 0.0006\\ 
13:27:59.91, -31:24:44.09 & 0.0405 $\pm$ 0.004 & 13:27:44.40, -31:28:01.46 & 0.0056 $\pm$ 0.0006\\ 
13:28:08.04, -31:30:23.77 & 0.0404 $\pm$ 0.004 & 13:27:57.50, -31:31:48.11 & 0.0048 $\pm$ 0.0005\\ 
13:28:01.60, -31:28:17.03 & 0.038 $\pm$ 0.0038 & 13:27:59.04, -31:33:09.20 & 0.0043 $\pm$ 0.0004\\ 
13:27:58.27, -31:33:33.20 & 0.0374 $\pm$ 0.0037 & 13:27:58.70, -31:31:11.48 & 0.004 $\pm$ 0.0004\\ 
13:27:57.48, -31:33:34.21 & 0.037 $\pm$ 0.0037 & 13:27:58.65, -31:32:02.00 & 0.0027 $\pm$ 0.0003\\ 
13:28:14.22, -31:31:08.56 & 0.0325 $\pm$ 0.0032 & 13:27:52.09, -31:25:51.25 & 0.0025 $\pm$ 0.0002\\ 
13:28:01.00, -31:30:55.99 & 0.0313 $\pm$ 0.0031 & 13:28:00.15, -31:28:18.02 & 0.0017 $\pm$ 0.0002\\ 
13:27:50.44, -31:28:19.46 & 0.0292 $\pm$ 0.0029 & 13:28:01.09, -31:31:43.99 & 0.0006 $\pm$ 0.0001\\
\hline
\end{tabular}
\label{tab:subtracted_srcs}
\end{table*}

\end{document}